\documentclass{aip-cp}
\usepackage{etoolbox}
\makeatletter
\patchcmd{\@@tablenote}{\xdef}{\protected@xdef}{}{}
\makeatother

\usepackage[numbers]{natbib}
\usepackage{rotating}
\usepackage{graphicx}

\graphicspath{{art/}}

\begin{document}

\title{Composite Entanglement Topology and Extensional Rheology of Symmetric Ring-Linear Polymer Blends}

\author[aff1,aff2]{Thomas C. O'Connor\corref{cor1}}
\author[aff3]{Ting Ge}
\author[aff1]{Gary S. Grest}

\affil[aff1]{Sandia National Laboratories, Albuquerque, NM 87815, United States}
\affil[aff2]{Department of Materials Science and Engineering,
Carnegie Mellon University, Pittsburgh, PA
15213, United States}
\affil[aff3]{Department of Chemistry and Biochemistry,
University of South Carolina, Columbia, SC
29208, United States}

\corresp[cor1]{Corresponding author: thomaso@andrew.cmu.edu}

\maketitle

\begin{abstract}
Extensive molecular simulations are applied to characterize the equilibrium dynamics, entanglement topology, and nonlinear extensional rheology of symmetric ring-linear polymer blends with systematically varied ring fraction $\phi_R$.
Chains with degree of entanglement $Z\approx14$ mixed to produce 10 well-entangled systems with $\phi_R$ varying from neat linear to neat ring melts.
Primitive path analysis are used to visualize and quantify the structure of the composite ring-linear entanglement network.
We directly measure the quantity of ring-linear threading and linear-linear entanglement as a function of $\phi_R$, and identify with simple arguments a ring fraction $\phi_R\approx0.4$ where the topological constraints of the entanglement network are maximized.  
These topological analyses are used to rationalize the $\phi_R$-dependence of ring and linear chain dynamics, conformations, and blend viscosity.
Complimentary simulations of startup uniaxial elongation flows demonstrate the extensional stress overshoot observed in recent filament stretching experiments, and characterize how it depends on the blend composition and entanglement topology.
The overshoot is driven by an overstretching and recoil of ring polymer conformations that is caused by the convective unthreading of rings from linear chains.
This produces significant changes in the entanglement structure of blends that we directly visualize and quantify with primitive path analyses during flow. 

\end{abstract}

\section{INTRODUCTION}
Molten polymers realize a wide range of time-dependent mechanical and rheological properties that emerge from the entanglement of intermingling chains \cite{DeGennes1979,Doi1988}.
The structure of the collective entanglement network and how it impacts the diffusion and relaxation of individual polymers depends strongly on the specific chain architecture \cite{Ward2012mechanical,Dealy2018}.
Industries have long leveraged polymer architecture as a powerful \textit{physical} route for tuning polymer properties and processability, independent of monomer chemistry.
In tandem, researchers have developed sophisticated theoretical models relating linear and nonlinear polymer rheology to the entanglement structure and dynamics of linear \cite{Likhtman2002,Graham2003,Schieber2014,Katzarova2015a,Shchetnikava2019}, ring \cite{Rubinstein1986,Milner2010,Ge2016}, star \cite{Frischknecht2000,Frischknecht2002},  and more highly branched structures \cite{McLeish1999,Das2006}.
An even wider variety of rheological behaviors can be realized by blending polymers of  different architectures.
In such ``architectural blends'' chains of different species contribute to and diffuse within the intermolecular entanglement network differently, giving rise to new rheological time and length scales that are difficult to anticipate \textit{a priori} \cite{Dealy2018}.
While some progress has been made, especially for blends of linear and branched polymers \cite{Milner1998,Das2006}, many aspects of these structurally and dynamically complex liquids remains poorly understood.

The dynamics of entangled ring polymers are perhaps the least well understood among chain architectures due to their closed-loop topology which produces a linear viscoelastic behavior distinct from that of conventional linear polymers \cite{Kapnistos2008,Milner2010,Ge2016}.
Unlike linear chains, rings do not form a system spanning topological entanglement network.
Instead, the topological constraints of other chains cause rings to adopt fractal loop conformations with self-similar conformations in space and relaxation dynamics in time.
This replaces the rubbery plateau $G_e$ in the stress relaxation modulus $G_{L}(t)$ for entangled linear polymers by a power-law stress relaxation $G_e \left(t/\tau_e\right)^{-\alpha}$ in entangled ring polymers where the exponent $\alpha$ is slightly below $1/2$ for unentangled Rouse dynamics \cite{Kapnistos2008,Ge2016}.

Blending rings with linear chains produces even more intriguing viscoelastic behaviors due to the interplay of their distinct topologies.
The closed-loop structure of rings enables them to be threaded by other chains \cite{Mills1987,Yang2010,halverson12,Tsalikis2014a,Lee2015,Smrek2019,Zhou2019,Michieletto2021,Hagita2021}, which can significantly alter chain conformations \cite{Iyer2007,Subramanian2008,Jeong2017}, segmental dynamics \cite{Mills1987,Ge2016,Kruteva2017, Wang2020a}, and blend viscosity \cite{McKenna1989,Nam2009,halverson12,Peddireddy2020}.
Roovers showed for ring-linear blends of polybutadiene
with approximately 15.3 entanglements per chain that there was a maximum in zero-shear viscosity for a ring fraction  $\phi_R \sim 0.4$ \cite{Roovers1988}.
Kapnistos \textit{et al.} \cite{Kapnistos2008} systematically studied the effect of a low volume fraction $\phi_{L}$ of linear polymers on the stress relaxation modulus $G_{R}(t)$ entangled ring polymers were studied. The experiments showed that $G_{R}(t)$ is enhanced at $\phi_{L}$ almost two decades below $\phi_{L}^{*}$ at which the linear chains begin to overlap. Such an enhancement is attributed to the emergence of a transient network of the isolated linear chains bridged by ring-linear threadings.
Molecular dynamics simulations by Halverson \textit{et al.} \cite{Halverson2012} demonstrated that the added linear polymers increase the zero-shear viscosity of a ring polymer melt by $10\%$ at $\phi_{L} \approx \phi_{L}^{*}/5$.
Further, Halverson \textit{et al.} observed a non-monotonic dependence of the zero-shear viscosity on blend ration, with a maximum near $\phi_L\sim 0.5$.
This was recently demonstrated experimentally by Peddireddy \textit{et al.} with optical tweezers microrheology measurements on entangled solutions of ring and linear DNA cite{Peddireddy2020}.

More recently, Parisi \textit{et al.} \cite{Parisi2020} reported the linear viscoelasticity of ring-linear polymer blends at ring volume fraction $\phi_{R}$ below $\phi_{R}^{e}$ for the onset of entanglements between ring polymers. Their experiments and theoretical modeling show that the rings are trapped by the entanglements with linear chains and the associated stress relaxation is completed by the constraint release (CR) induced by the mobility of the linear chains. The model approximates the overall stress relaxation modulus as a linear combination of the contributions from the linear chains and rings 
\begin{equation}
    G(t) = (1-\phi_{R})G_{L}(t) + \phi_{R}G_{R}(t)
    \label{eq:eq1}
\end{equation}
$G_{L}(t)$ includes a fast Rouse relaxation $G_{F, Rouse}(t)$ for $t<\tau_e$, a longitudinal relaxation $G_{L}(t)$ along the confining tube for $\tau_e < t < \tau_R$, where $\tau_R$ is the Rouse time of the chain, and the entanglement-involved relaxation $G_e \mu(t)R(t)$, where the single-chain repation and contour length fluctuation (CLF) and the multi-chain CR are described by $\mu(t)$ and $R(t)$, respectively. The same $G_{F, Rouse}(t)$ and $G_{L}(t)$ are used in the expression of $G_{R}(t)$, while the entanglement-involved relaxation $G_{CR}(t)$ is determined by a pure CR process that is controlled by the repation of the linear chains. Integrating the theoretical expression of $G(t)$ gives that the zero-shear viscosity of the blend $\eta = \int_{0}^{\infty} G(t)dt$ is above the pure linear melt viscosity $\eta_{L}$ and increases linearly with $\phi_{R}$ The model describes well the experimental $G(t)$ and $\eta$ or different $\phi_{R} < \phi_{R}^{e}$. An alternative model for $G(t)$ of ring-linear blends at low $\phi_{R}$ is developed by Hou. The same approximation in Eq.~\ref{eq:eq1} is used but with several modifications of $G_{L}(t)$ and $G_{R}(t)$. For $G_{L}(t)$, the traditional Likhtman-McLeish expression is used.\cite{Likhtman2002} For $G_{F,Rouse}(t)$ and $G_L(t)$ in $G_{R}(t)$, while Parisi \textit{et al} \cite{Parisi2020} simply used the expressions for linear polymers. Hou \cite{Hou2020} derived two new expressions for rings. For $G_{CR}(t)$ in $G_{R}(t)$, Hou used the Rouse description originally developed by de Gennes instead of the self-consistent description by Rubinstein and Colby. Both the models by Parisi \textit{et al}. and Hou can fit the experimental data of $G(t)$ and thus $\eta$ very well, although differences appear for long times beyond the time range in the experiments.
Threading of rings has also been shown to significantly alter the dynamics of chains during nonlinear shear and extensional flows \cite{Huang2019,Zhou2019}.
Using filament stretching rheometry  Huang \textit{et al.}, showed that entangled melts of polystyrene rings are highly sensitive to extensional flows, exhibiting massive extension-rate thickening even when the Weissenberg number was much smaller than the ring relaxation time.
O'Connor \textit{et al.} used nonequilibrium molecular simulations to demonstrate that this behavior was due to flow driving rings to \textit{self-thread} and link together to form supramolecular daisy-chains.\cite{OConnor2020}
In ring-linear blends, nonlinear flows can also drive the unthreading of rings from linear chains through convective constraint release.
This has been directly observed in companion single-molecule experiments \cite{Zhou2019,Zhou2021} and Brownian dynamics simulations \cite{Young2020} of semidilute ring-linear DNA solutions undergoing planar extensional flow.
These studies observed large fluctuations in ring elongation in flow, as well as ring overstretching and recoil, driven by transient ring-linear threadings. 
Borger \textit{et al.} identified a similar unthreading mechanism to explain the observed overshoot in extensional stress $\sigma_E$ during the uniaxial elongation of ring-linear blends of polystyrene \cite{Borger2020}.
They combined molecular simulations, filament stretching rheometry, and SANS experiments to show  that the overshoot could be explained by the flow-driven unthreading of rings.

\begin{figure}[h!]
  \centerline{\includegraphics[width=0.75\textwidth]{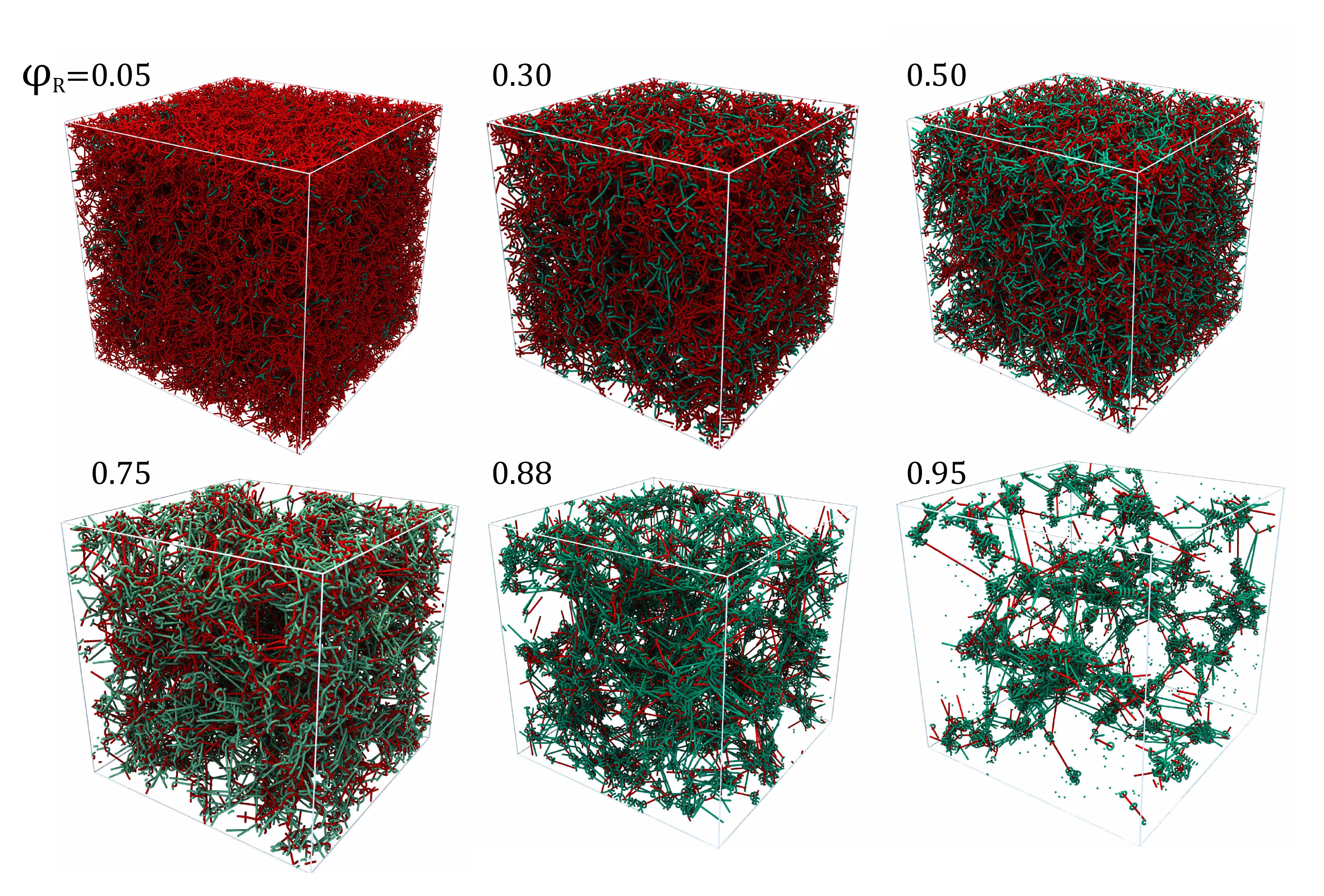}}
  \caption{
  Blend network topologies generated by primitive path analysis for ring-linear blends at six different $\phi_R$. 
  Primitive paths of linear and ring chains are colored red and green, respectively.
  Linear chain ends are held fixed during the PPA while rings are unconstrained, such that ring primitive paths collapse to points unless they are threaded by other chains.
  }
  \label{fig:networks}
\end{figure}

These prior works show the central role of ring-linear in the topological structure and dynamics of ring-linear blends.
For small amounts of ``architectural contamination,'' the linear viscoelastic models of Parisi 
textit{et al.} \cite{Parisi2020} and Hou \cite{Hou2020} provide limiting pictures for how threadings modify neat melt rheology. 
However, it is difficult to know how to extend these ideas to compositions $\phi_R$ not near a neat melt.  
For arbitrary $\phi_R$, the structure of topological constraints within blends become quite complex and difficult to predict a priori, but molecular simulations allow us to directly can characterize these complex networks with methods like the primitive path analysis  \cite{Everaers2004,Sukumaran2005},  as shown in Figure \ref{fig:networks}.
In these complex topological states, both linear and ring polymers entangle with their own species as well as with their counterparts. As a result, the entangled dynamics of both linear chains and rings are qualitatively changed with respect to their dynamics in a pure polymer matrix. 

In this study we present extensive molecular simulation results characterizing the structure, topology, and dynamics of symmetric ring-linear blends in equilibrium and undergoing uniaxial elongational flow.
We characterize the conformational statistics, diffusive dynamics, and primitive path topologies of well-entangled ($Z\approx14)$ ring and linear chains in their neat melts and eight ring-linear blends with $\phi_R$ varying from 0.05 to 0.95.
Uniaxial elongation flow simulations show the development and $\phi_R$-dependence of the extensional stress overshoot observed by Borger \textit{et al.},\cite{Borger2020} and also track how the structural and topological statistics of the blends evolve during flow.
These analysis enable us to directly quantify and visualize the flow-driven unthreading of rings from the linear network during elongation.
While prior simulation studies have separately characterized some of these quantities for different ring-linear systems \cite{Tsalikis2014,Lee2015,Jeong2017,Young2020,Wang2020a}, these results are difficult to aggregate due to variations and limitations in system size, concentration, polymer size, and the degree of ring-linear threading.
Thus, in order to build a fuller and clearer picture of topological threading in ring-linear blends, we have generated systems of well-entangled chains with system sizes that are a factor 5-10 times larger than typically found in the literature. 
These large systems allow us to resolve the structural and dynamic statistics of both chain architectures at all $\phi_R$.
In addition, by aggregating both equilibrium and non-equilibrium data for this ensemble of blends, we are able to draw connections between trends in chain dynamics, melt topology, and rheological behavior.

\section{MODEL AND SIMULATION METHODOLOGY}

We model polymer chains with a semi-flexible bead-spring model. Monomers have spherical symmetry and interact through a purely repulsive Lennard-Jones potential that is truncated at $r_c=2^{1/6} \sigma$ \cite{Kremer1990}. Chains of $N$ monomers are formed by connecting monomers with Finitely-Extensible Nonlinear Elastic (FENE) bonds with standard parameters \cite{Kremer1990,Hsu2016}. A bending stiffness $U_\theta = k_\theta(1+\cos\theta)$ with $k_\theta = 1.5 u$, is applied between pairs of consecutive bonds with an angle $\theta$ relative to parallel orientation. Throughout this paper, all physical quantities are expressed in Lennard-Jones units of monomer mass $m$, length $\sigma$, energy $u$, and time $\tau=\sqrt{m\sigma^2/u}$. All simulations are performed
using LAMMPS \cite{Plimpton1995} with a time step $\Delta t = 0.01$ $\tau$ for the equilibrium runs and 0.007 $\tau$ for the elongational runs.

The dynamics of entangled linear polymers is dominated by the entanglement segment length $N_e$, that emerges as chains intermingle and topological constrain each other's diffusion. The rheology of linear chains of length $N$ can be fully characterized by three numbers: the degree of entanglement $Z=N/N_e$, and a chemically specific entanglement time $\tau_e\sim N_e^2$ and plateau modulus $G_e\sim\rho k_{\rm B} T/N_e$ \cite{Doi1988}. $G_e$ is the free energy density of the entanglement network formed by linear chain entanglements, and $k_{\rm B}$ is Boltzmann's constant. 
The linear viscoelastic parameters of this bead-spring model have been thoroughly characterized in prior studies \cite{Hsu2016,Moreira2015,Parisi2021}, yielding $\tau_e\approx 1.98 \times 10^3 \tau$ and $N_e= 28\pm2$ beads \cite{Sukumaran2005,Hoy2009}. For $N=400$, there are $Z=N/N_e\approx14$ rheological entanglements per chain in a linear melt.

In unconcatenated ring melts, chains do not develop a long-range entanglement network. Instead, the topological constraints of other chains cause individual rings to form fractal globule conformations of double-folded and interpenetrating loops \cite{Smrek2019}. The length of the fundamental loop segments $N_\ell$ should be $\sim N_e$, and larger ring segments form fractal loops composed out of these fundamental loops. In this picture, the relaxation of large loops is due to a self-similar cascade of smaller loop relaxations that produces a distinct power-law relaxation modulus $G(t)\sim t^{\beta}$, with $\beta=-3/7$. 

It is important to note, that while this fundamental loop size is expected to be of the same order as $N_e$, there is no reason to presume that the two should be the same. Indeed, in their systematic studies of bead-spring ring melts, Ge et al. \cite{Ge2016} found a fundamental loop size and relaxation time close to but not equal to the $N_e$ and $\tau_e$ measured for linear systems. In this paper, we apply the linear viscoelastic theory identified and fit by Ge et al. for pure ring melts of the same bead-spring model.

All ring-linear blend samples were generated based on two samples of $N=400$ from Halverson et al.  \cite{halverson12}. In one sample, the number of rings $M_L=113$ and the number of linear chains $M_L=113$, corresponding to ring volume fraction $\phi_R=0.50$. In the second sample, $M_R=200$ and $M_L=26$, corresponding to $\phi_R=0.88$. Larger samples of each were made by replicating the original samples by a factor $f=2$ along each of the three orthogonal directions. The new samples of $\phi_R=$ 0.05, 0.10, 0.20, 0.30, and 0.75 were made by first removing a fraction of either ring or linear chains from the original $\phi_R=0.50$ system and then replicating the sample with new $\phi_R$ to a larger size. $f=3$ for the replication of the sample with $\phi_R=0.05$, while $f=2$ for the replication of the other samples. The new sample of $\phi_R=0.95$ was made by first removing a fraction of linear chains from the original sample of $\phi_R=0.88$ and then replicating the sample with $\phi_R=0.95$ along each direction by $f=2$. All the newly generated samples were equilibrated first at constant temperature $T=u/k_B$ using a Nose-Hoover thermostat with a characteristic damping time 100$\tau$ at constant pressure $P=5.0 u/\sigma^3$ using a Nose-Hoover barostat with a characteristic damping time 100$\tau$. This stage of the equilibration was run for up to 20,000$\tau$, which was sufficient to allow the density $\rho$ to stabilize around $0.85 \sigma^{-3}$. Subsequently, all the samples were run at fixed volume and $T=u/k_B$ using Nose-Hoover thermostat with a characteristic damping time $100\tau$. Each system was then run between $3.0 \times 10^7$ to $1.0 \times 10^8\tau$. 
Table 1 lists $\phi_R$, $M_R$, $M_L$, the simulation box size $L$, run time $t_{\rm run}$, the mean squared radius of gyration of the ring and linear chains and the terminal relaxation time $\tau_d$ for the mean squared displacement of a monomer in the ring and linear chains equals to $3<R_g^2>$ \cite{Hsu2016} for each system studied.  
For the rings, the mean squared displacement of a monomer was averaged over all the beads in the chain, while for linear chains, the mean squared displacement was averaged over the inner 5 beads.

The stress relaxation modulus $G(t)$ was measured for each system using the Green-Kubo relation $G(t) = (V/k_BT) \left\langle\sigma_{\alpha\beta}(t)\sigma_{\alpha\beta}(0)\right\rangle$ where $\sigma_{\alpha\beta}$ are the off-diagonal components $xy$, $xz$ and $yz$ of the stress. In order to reduce the noise in $G(t)$, the stress autocorrelation function is calculated using pre-averaged stresses $\bar\sigma_{\alpha\beta}$ averaged over 1000 steps. Results for $G(t)$ presented here are an average of the 3 components.

We impose uniaxial extensional flows by applying an affine stretch $\Lambda(t)$ along the z-axis of our simulation cell at a constant Hencky strain-rate $\varepsilon=\partial\Lambda/\partial{t}$. This stretches the system exponentially in time. Since the fluid is nearly incompressible, both the x and y dimensions of the system contract as $\Lambda(t)^{-1/2}$ to preserve volume. Generalized Kraynik-Reinelt (GKR) boundary conditions are employed during flow to systematically remap the simulation box and prevent it from becoming too skewed.\cite{Dobson2014,Nicholson2016,OConnor2018PRL}

A primitive path analysis (PPA) \cite{Everaers2004,Sukumaran2005} is performed to determine the nature and extent of ring-linear entanglements in the blend. The ends of the linear chains are fixed in space while no constraints are applied to the ring polymers \cite{Subramanian2008}. The angular and nonbonded intrachain interactions are switched off and the temperature lowered to $0.001 \epsilon/k_{\rm B}$. The MD simulation is then run for $2000 \tau$ which is sufficient to reach equilibrium. This drives linear chains to contract to their primitive paths, constrained by entanglements, while rings collapse to points, unless they are threaded and constrained by linear chains. In the pure ring melt, the rings do not form conventional entanglements with other rings and the rings collapse to points during PPA. However, rings in the blend do not collapse to points due to the presence of the linear chains threading the rings.  From the PPA, the contour length of primitive-paths $L_{pp}$ for both linear and ring chains is measured. The number of linear chains threading a ring is determined by counting the number of linear chains which have at least one bead that is one entanglement length $N_e =28$ \cite{Sukumaran2005,Hoy2009} from the end of the chain in contact with a ring after the PPA is completed.

\section{Results}
\subsection{Equilibrium Structure and Dynamics}

\begin{figure}[h!]
  \centerline{\includegraphics[width=0.5\textwidth]{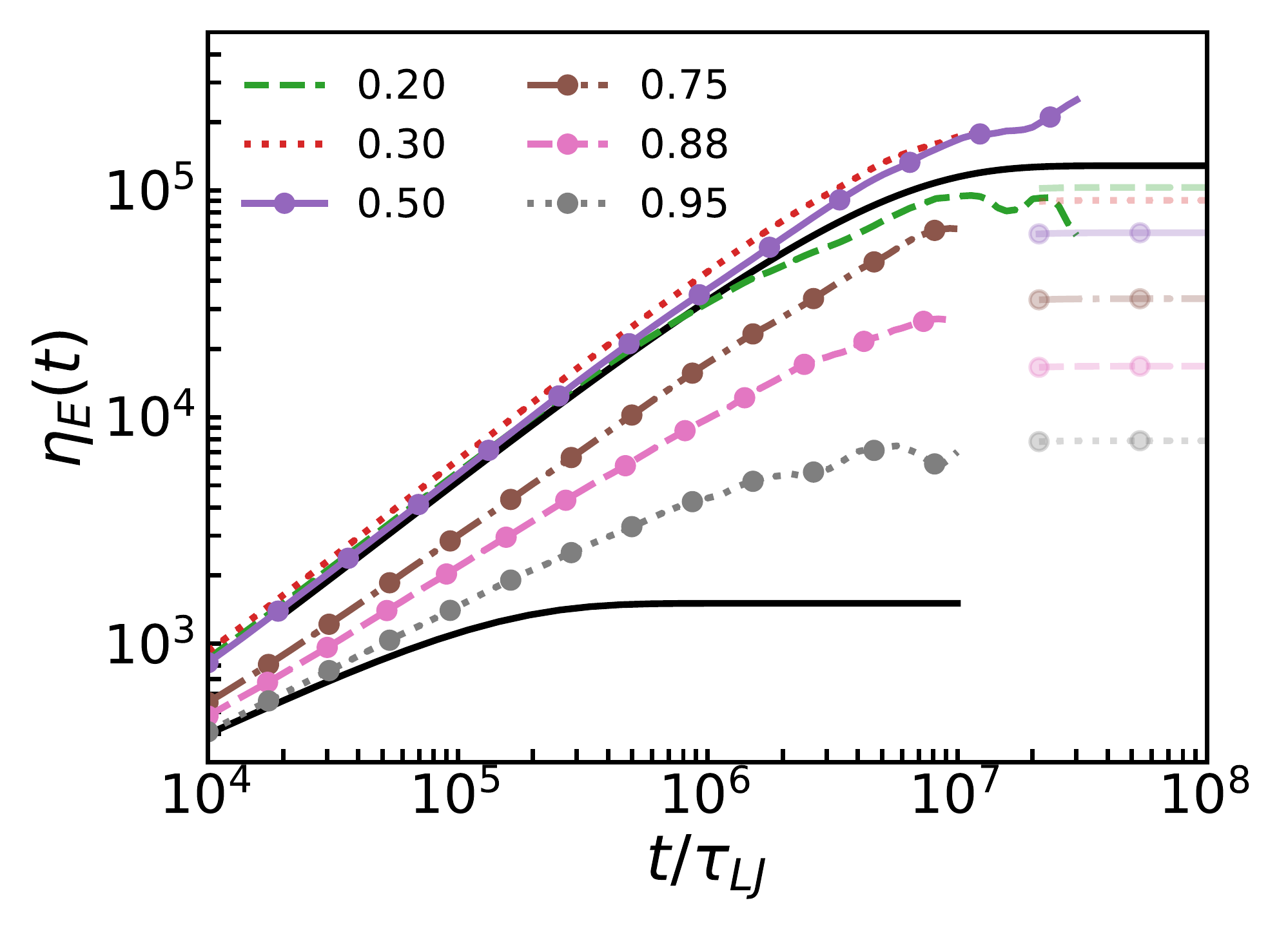}}
  \caption{
  Linear viscoelastic envelopes (LVE) for six ring-linear blends, computed by Green-Kubo analysis of equilibrium melts.
  The two solid-black curves are analytic LVEs for neat ring and linear melts, derived from Likhtman-McLeish and Fractal-Loopy-Globule theories, respectively.\cite{Likhtman2002,Ge2016}
  Faded horizontal lines are Newtonian plateaus for an ideal mixing of the two neat LVEs, which tend to under predict the blend viscosity at most $\phi_R$.
  }
  \label{fig:lve}
\end{figure}

Our focus in this study will be characterizing how the rheology, structure, and dynamics of entangled blends change as the ring fraction $\phi_R$ is systematically varied. The ring fractions studied are listed in Table 1. We begin by directly measuring the mean squared displacement of the chains and the linear viscoelastic stress relaxation modulus for all $\phi_R$ except our lowest $\phi_R=0.05$. This system is excluded because of its prohibitively large size, which was necessary to have enough rings for meaningful statistics, to run long enough to extract dynamic quantities.  The linear viscoelastic envelope $\eta(t)=\int_0^t G(t) dt$ for blends of different $\phi_R$ (colors) is shown in Figure \ref{fig:lve} and compared to the LVE curves for pure melts (solid black). The pure melt LVEs for linear and ring melts are computed from the theory by Likhtman-McLeish \cite{Likhtman2002} and the FLG model of Ge et al.~\cite{Ge2016} respectively.  Direct calculation of $\eta(t)$ is computationally costly, and the statistics of our curves at large $t$ are not sufficient to make strong quantitative statements. However, we can make some useful qualitative observations.

Most notably, our data show that blending ring and linear chains does not produce a simple mixing of the linear response of the neat melts, consistent with recent experiments.\cite{Peddireddy2020} A simple mixing would produce a linear interpolation between the neat melt LVEs and a linear decrease in the Newtonian viscosity with increasing $\phi_R$. The expected values $\eta_0(\phi)$ for this simple case are shown by the faded horizontal lines in Figure \ref{fig:lve}. However, our data clearly show that $\eta_0(\phi)$ is nonmonotonic and exhibits a maximum. Indeed, $\eta_0(\phi)$ increases for a small number of rings in a linear melt - $\phi_R$ near 0 – and for a small number of linear chains in a ring melt - $\phi_R$ near 1. This is consistent with prior studies that have explored the rheology of both limits of lightly contaminated blends.

Different pictures have been proposed to understand the increase in $\eta_0$ in the two limits of $\phi_R$. For $\phi_R$ near 0, one can imagine a small number of rings threaded by linear chains so that they are ``sewn’’ into a fully developed linear entanglement network. The ring dynamics are strongly suppressed by the entanglement network, because a ring threaded by linear chains can only diffuse as all the linear chains release topological constraints by reptating. Thus, the embedded rings will have a \textit{longer} relaxation time than the linear chains, producing an enhancement of the viscosity. This is confirmed by the increase in terminal relaxation time $\tau_d^R$ as $\phi_R$ decreases as seen from the results in Table \ref{tab:eql}. At the other extreme there are few linear chains within a ring melt. The linear chains extend much larger distances than the compact globular rings as seen in Table 1 and they thread multiple rings. These additional threading constraints also slow ring diffusion relative to the neat melt and, as we will show, facilitate the development of a composite entanglement network of linear-ring-linear couplings.

\begin{table*}
\begin{center}
\caption{Ring fraction $\phi_R$, number of rings $M_R$, number of linear chains $M_L$, the simulation box size $L$, run time $t_{\rm run}$, the mean squared radius of gyration of the ring $\langle R^2_{gR}\rangle/\sigma$ and linear chains $\langle R^2_{gL}\rangle/\sigma$ and the terminal relaxation time for the rings $\tau_d^R$ and linear chains $\tau_d^L$}
\begin{tabular}{|c|c|c|c|c|c|c|c|c|}
  \hline
 $\phi_R$  & $M_R$ & $M_L$ &$ L/\sigma$ & $t_{\rm run}/\tau$&  $\langle R^2_{gR}\rangle/\sigma$ & $\langle R^2_{gL}\rangle/\sigma^2$& $\tau_d^R/\tau$ &$\tau_d^L/\tau$ \\
  \hline
  1.0  & 1600& - & 45.5 & $3.0 \times 10^7$ & 53.1 & - & $1.6 \times 10^5$ & - \\
 0.95  & 1600 & 80 & 92.5 & $4.1 \times 10^7$ & 53.6 & 188.3 & $2.4\times10^5$ & $9.8\times10^6$\\
 0.88 & 1600 & 208 & 128.6 & $4.0 \times 10^7$ & 54.7 & 188.5 & $6.3\times10^5$ & $1.3\times10^7$ \\
 0.75 & 904 & 304 & 112.4 & $6.0 \times 10^7$ & 57.1 & 184.0 & $1.7\times10^6$ &  $1.5\times10^7$\\
0.50  & 904 & 904 & 128.6 & $8.1 \times 10^7$ & 64.4 & 181.7 & $1.4\times10^7$ & $1.5\times10^5$ \\
0.30  & 384 & 904 & 114.9 & $8.6 \times 10^7$ & 72.5 & 179.4 & $2.3\times10^7$ & $1.5\times10^7$\\
0.20  & 224 & 904 & 110.0 & $1.0 \times 10^8$ & 77.2 & 178.8 & $2.6\times10^7$ & $1.5\times10^7$\\
0.10  & 124 & 904 & 106.5 & $7.0 \times 10^7$ & 81.9 & 178.5 & $2.8\times10^7$ & $1.5\times10^7$\\
0.05  & 162 & 3051 & 114.8 & $3.0 \times 10^7$ & 82.9 & 180.8 & - & -\\
\hline
\end{tabular}
\label{tab:eql}
\end{center}
\end{table*}

Figure \ref{fig:networks} shows snapshots of blend network structures produced by PPA for six values of $\phi_R$. Our PPA method shrinks the chain contours of linear chains (red) while holding linear chain ends fixed in space. Rings are left unconstrained. This produces configurations that reveal the hybrid network of linear (red) and ring (green) entanglements and threadings in the blends. Notably, PPA drives \textit{unthreaded} rings to collapse into points as seen for $\phi_R=0.95$ in Figure \ref{fig:networks}. However, the topological constraints of threaded rings produce open structures after PPA. This allows us to characterize the fraction of threaded rings and how many linear chains thread them for various $\phi_R$.

Figure \ref{fig:networks} illustrates explicitly both the limits of few rings heavily threaded into a linear entanglement network ($\phi_{R}=0.05$) and the other extreme of few linear chains threading many rings ($\phi_R=0.95$). The configurations also illustrate a qualitative change in the topological structure and constraint density as linear chains are replaced by rings. For $\phi_R$ near zero the topology is dominated by the dense network of linear chain entanglements (red chains). Replacing linear chains with rings removes linear-linear entanglement constraints and replaces them with new constraints due to ring-linear threadings. This exchange produces an intricate hybrid network topology of linear (red) and ring (green) chains that is particularly notable for $\phi_R=0.5$ and $0.75$ in Figure \ref{fig:networks}. Eventually the network must become sparse and break down as $\phi_R$ approaches 1.0 since pure rings do not form persistent entanglements in equilibrium without ring-linear threadings to mediate them. Thus, the network must rapidly break down as the fraction of linear chains, and number of ring-linear threadings, becomes small. 

\begin{figure}[h!]
  \centerline{\includegraphics[width=0.75\textwidth]{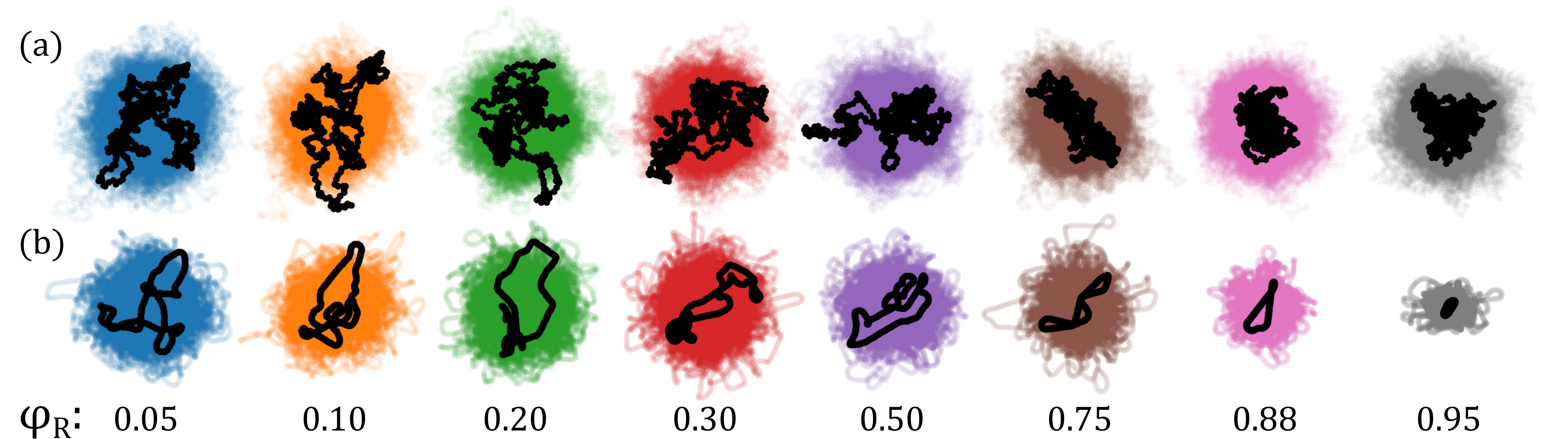}}
  \caption{
  Equilibrium ensembles of (a) ring conformations and (b) primitive paths from blends at different $\phi_R$.
  For each ensemble, all chains are made translucent and shifted to a common center of mass, producing a colored cloud. 
  The black chain emphasized in each ensemble is the ring with $\langle R_{gR}^2\rangle$ or $\langle L_{pp}^{R2}\rangle$ closest to the root-mean-square value.
  }
  \label{fig:eqlconf}
\end{figure}

Ring-linear threadings enable ring polymers to topologically entangle with other rings and with the entanglement network of linear chains.
This produces large changes in the equilibrium dynamics and conformations of ring polymers.
Figure \ref{fig:eqlconf}(a) and (b) show ensembles of equilibrium ring conformations and ring primitive paths generated by PPA for blends at each $\phi_R$, respectively. 
The colored clouds in Figure \ref{fig:eqlconf} are generated by shifting all rings in the system to a common center of mass.
The black highlighted chain in each ensemble is the chain with a gyration radius $R_g^R=\langle R_{gR}^{2}\rangle^{1/2}$ or ring primitive path length $L_{pp}^R=\langle L_{pp}^{R2}\rangle^{1/2}$ closest to the ensemble average.
Considering both panels, we see that rings are most compact in pure melts and their conformations become larger and more open as $\phi_R$ decreases and the fraction of linear chains increases.
This opening of ring conformations directly corresponds with an increase in the ring primitive path due to linear threading.
We note, melts of pure rings will collapse to points during PPA, producing $L_{pp}^R\approx0$. Thus, $L_{pp}^R$ can only be greater than zero due to linear threading of rings.

\begin{figure}[h!]
  \centerline{\includegraphics[width=0.5\textwidth]{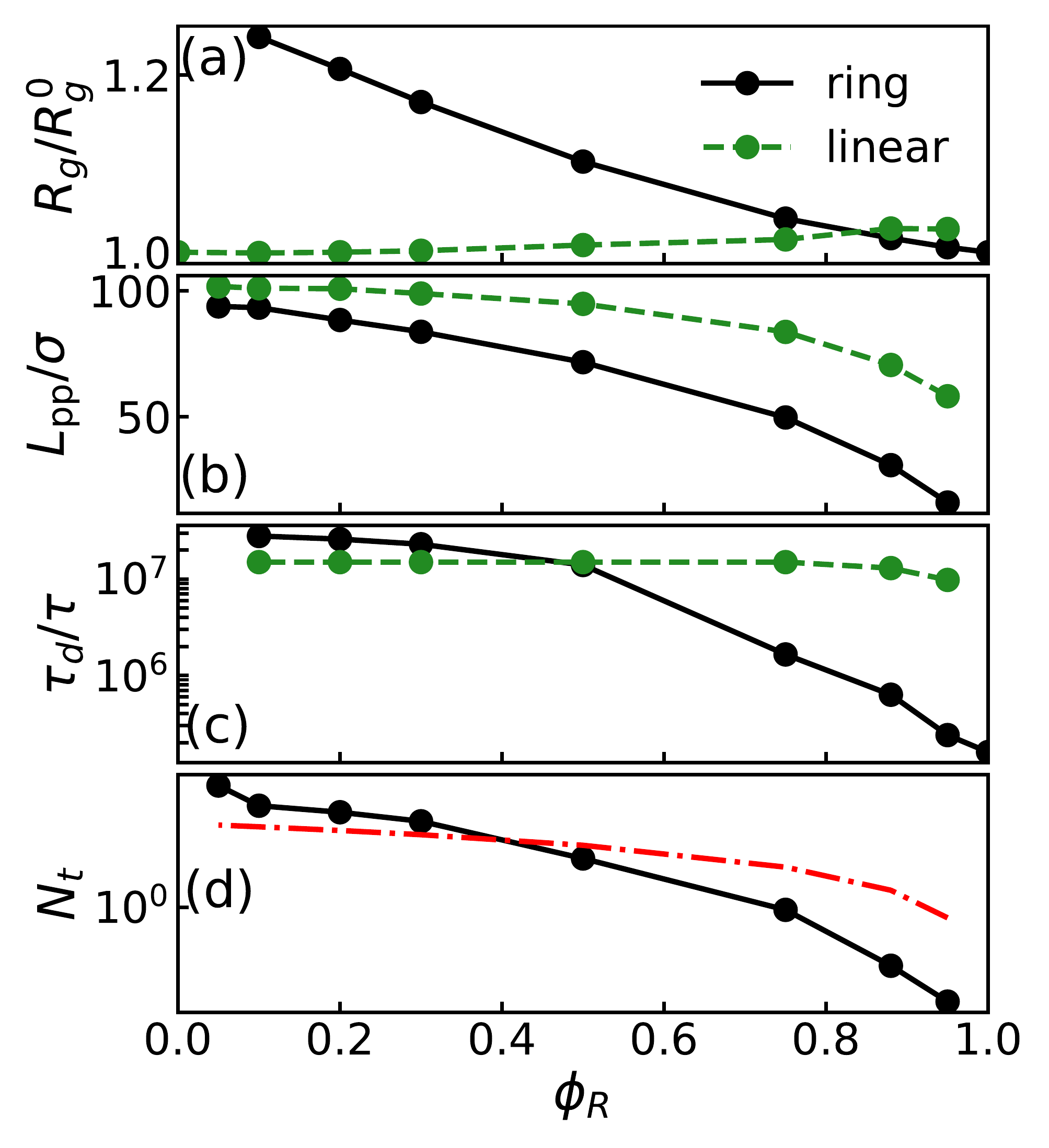}}
  \caption{
  Equilibrium blend statistics for ring (solid black) and linear (green-dashed) chains for varying $\phi_R$.
  (a) Change in the root mean squared radius of gyration $R_g$ relative to its value in a chain's corresponding pure melt $R_g^0$.
  (b) Root mean squared primitive path lengths $L_{pp}$. Ring $L_{pp}^R\rightarrow0$ at $\phi_R=1.0$.
  (c) The diffusive relaxation times $\tau_D$ are computed as the average time required for a chain to diffuse a distance equal to 3$R_g$.
  (d) Average number of linear threadings per ring measured by primitive path analysis. The red dot-dash line is the curve $(1-\phi_R)Z$, the average number of linear-linear entanglements a linear chain is expected to form when diluted in rings.
  Rings contribute fewer topological constraints to the blend network than linear chains when $N_t<(1-\phi_R)Z$.
  }
  \label{fig:eqlstats}
\end{figure}

These trends with $\phi_R$ are plotted for $R_g/R_g^0$ and $L_{pp}$ for both ring and linear chains in Fig.~\ref{fig:eqlstats}(a) and (b). Results for $\langle R_g^2\rangle$ are given in Table 1.  Here, $R_g^0$ is the equilibrium radius of gyration for either species in their corresponding \textit{neat} melts.
Linear chains show almost no change in $R_g$, swelling by only a few percent for the largest values of $\phi_R$.
Clearly the linear chains experience segments of neighboring rings much like they experience linear chain segments.
Otherwise we would expect a much larger change in $R_g$.
In contrast, rings exhibit significant swelling, with $R_g$ increasing by more than $20\%$ as $\phi_R$ decreases from 1.0 to 0.05, consistent with prior simulation \cite{Jeong2017} and experimental studies.\cite{Iwamoto2018}
These sensitivities are mirrored in the primitive path lengths $L_{pp}$ plotted in Fig.~\ref{fig:eqlstats} (b).
Linear primitive paths show little change at low $\phi_R$ but begin to decrease gradually once $\phi_R>0.5$. 
Ring $L_{pp}^R$ are comparable to linear chains for small $\phi_R$, but decrease dramatically with increasing $\phi_R$ and approach 0 at $\phi_R=1.0$.

The decrease in both linear and ring $L_{pp}$ with increasing $\phi_R$ are due to the removal of both linear entanglements and ring-linear threadings.
The average number of linear chains threading a ring $N_t$ is plotted for each $\phi_R$ in Fig.~\ref{fig:eqlstats}(d). 
As we observed in Fig.~\ref{fig:networks}, rings are highly embedded at small $\phi_R$, with $N_t\sim50$ threadings per ring.
This decreases rapidly to $N_t\sim1$ once $\phi_R\geq0.75$.
For large $N_t$, ring segments are well embedded in a linear entanglement network, and the ring entanglement segments adopt conformations similar to linear entanglements.
This produces similar $L_{pp}$ for both ring and linear chains at low $\phi_R$.
As $\phi_R\rightarrow 1.0$ and $N_t\rightarrow0$, $L_{pp}^R\rightarrow0$ for rings, and one might expect $L_{pp}^L\rightarrow R_{ee}$ for linear chains, where $R_{ee} = \langle R_{ee}^2\rangle^{1/2}$ is the root mean squared end-to-end distance. 
While $\langle R_{ee}^2\rangle^{1/2}\sim 33\sigma$ for the linear chains at all $\phi_R$, $L_{pp}^L$ remains significantly larger ($\sim58\sigma$) even at $\phi_R=0.95$.
The physical picture is illustrated by the $\phi_R=0.95$ configuration in Figure \ref{fig:networks}, which shows linear primitive paths composed of long segments connected by few ring junctions acting as ``slip-links'' between separated linear chains.
Thus, we come to appreciate that even as $\phi_R\rightarrow1$ and $N_t\rightarrow0$, such that each ring is threaded by 1 or fewer linear chains on average, each linear chain will still be threaded through many different rings, and some fraction of them are likely to be threaded by another linear chain.
This small population of doubly threaded rings is enough to constrain linear primitive paths such that $L_{pp}^L> \langle R_{ee}^2\rangle^{1/2}$.

The diffusive relaxation times $\tau_d$ for ring and linear chains at different $\phi_R$ are shown in Fig.~\ref{fig:eqlstats}(c).
Note, the $\tau_d$ axis is plotted on a logarithmic scale, because in their neat melts rings diffuse much faster than linear chains, with $\tau_d\sim N^{2.4}$ rather than a linear chains $\tau_d\sim N^{3.4}$.
This is due to neat rings being much more compact (Figure \ref{fig:eqlconf}), and not forming direct entanglements with each other.
Linear chain relaxation shows very little change with $\phi_R$. Reminiscent of the structural data for $R_g$ in Fig.~\ref{fig:eqlstats}(a), the linear chains appear dynamically insensitive to whether they are reptating through linear or ring segments.
This is likely because the chains are long enough ($Z\approx14$) that it is difficult for a reptating chain end to distinguish between segments of of a ring and segments of a linear chain. This may change if the two architectures had very different chain lengths.

In stark contrast to linear chains, ring dynamics slow dramatically as $\phi_R$ decreases and linear threadings are introduced.
Ring relaxation times increase by nearly two orders of magnitude between $\phi_R=0.95$ and $\phi_R=0.5$, at which point $\tau_D^R$ and $\tau_D^L$ for ring and linear chains, respectively, are nearly the same.
For even smaller $\phi_R$, $\tau_d^R$ continues to increase, becoming twice $\tau_d^L$ for $\phi_R=0.05$.
The fact that highly threaded rings diffuse slower than linear chains makes sense in the context of the observations of Parisi \textit{et al.}.\cite{Parisi2020}
Highly threaded rings are embedded in the linear entanglement network such that ring diffusion requires linear chains to release threading constraints by reptation, which occurs on times $\sim\tau_D^L$ for linear chains.
When $N_t$ becomes large, many linear threadings must be released for rings to diffuse, driving ring diffusion to be slower than linear diffusion.

The terminal relaxation time $\tau_d^R$ for rings begins to decrease rapidly once $\phi_R>0.5$.
This suggests that the topological structure of the blends begins to change rapidly in this regime of $\phi_R$.
As illustrated in Figure \ref{fig:networks}, this appears to be the dynamic signature of the deterioration of the ring-linear entanglement network as topological constraints are rapidly lost with increasing $\phi_R$.
We can estimate the $\phi_R$ where we expect this deterioration to begin by considering the net change in topological constraints as linear chains are replaced by rings.
To do so, we will consider two types of topological constraints in the blend networks: linear-linear entanglements and ring-linear threadings.
In a pure linear melt ($\phi_R=0$), each linear chain contributes $\sim Z\approx 14$ linear-linear entanglements.
In a blend with ring fraction $\phi_R$, the linear network is diluted such that each linear chain forms $\sim (1-\phi_R) Z$ linear-linear entanglements.\cite{Ruymbeke2014a}
At the same $\phi_R$, each ring contributes an average number $N_t$ of ring-linear threading constraints.
Roughly speaking, replacing a single linear chain with a single ring removes all of the linear chain's linear-linear entanglements and replaces them with the ring's $N_t$ threadings.

We expect the blend network to deteriorate with increasing $\phi_R$ when $N_t(\phi_R)<(1-\phi_R) Z$, or when rings contribute fewer threadings than linear chains contribute linear-linear entanglements.
The curve $(1-\phi_R) Z$ is plotted in Figure \ref{fig:eqlstats}(d) as red dot-dash line.
As can be seen, $N_t$ crosses below this curve for $\phi_R\approx 0.4$, which corresponds closely with the rapid drop in $\tau_D^R$ in Figure \ref{fig:eqlstats}(c) and the thinning of network constraints in the configurations in Figure \ref{fig:networks}.

\subsection{Stress Overshoot During Nonlinear Elongation}

The composite threading network of blends produces distinctive nonlinear dynamics and rheology not seen in pure linear or ring melts \cite{Halverson2012,Peddireddy2020,Borger2020,Young2020,Zhou2020}.
In particular, Borger \textit{et al.} \cite{Borger2020} have recently shown that a polystyrene blend with $\phi_R=0.3$ undergoing uniaxial elongation exhibits a prominent nonlinear overshoot in extensional stress and viscosity over a wide range of $Wi_R$.
Borger \textit{et al.} combined experiments and molecular simulations to conclude that the overshoot in extensional stress was due to the overstretching and unthreading of rings embedded in the linear entanglement network.
Heavily threaded rings relax slower than linear reptation and are forced to deform affinely with the linear network until their threading constraints are convectively released by the flow.
This unthreading transition allows rings to recoil producing a significant stress drop.

Borger \textit{et al.} focused on the extensional rheology of one symmetric blend with $\phi_R=0.3$ for a range of strain rates.
Here, we build on their analysis and characterize the development and mechanistic details of this stress overshoot during startup uniaxial elongation for blends of varying $\phi_R$.
We have simulated startup flow for a range of strain rates, but for the majority of what follows we limit our focus to the $\phi_R$ dependence of blends elongated at one strain rate $\dot{\varepsilon}_H=7.8\times10^{-6}\tau_{LJ}^{-1}$. 
This rate corresponds to a Weissenberg number $Wi=\dot{\varepsilon}_H \tau_D^R\approx1.1$ for rings in the pure ring melt, and is strong enough to produce significant nonlinear response.\cite{Ge2016,OConnor2020,Borger2020}
We do this to both simplify our discussion, and because we observe that the $\phi_R$-dependent features of the overshoot are relatively insensitive to the choice of strain rate.
We will touch on these similarities with rate dependent data as we go.

\begin{figure}[h!]
  \centerline{\includegraphics[width=0.5\textwidth]{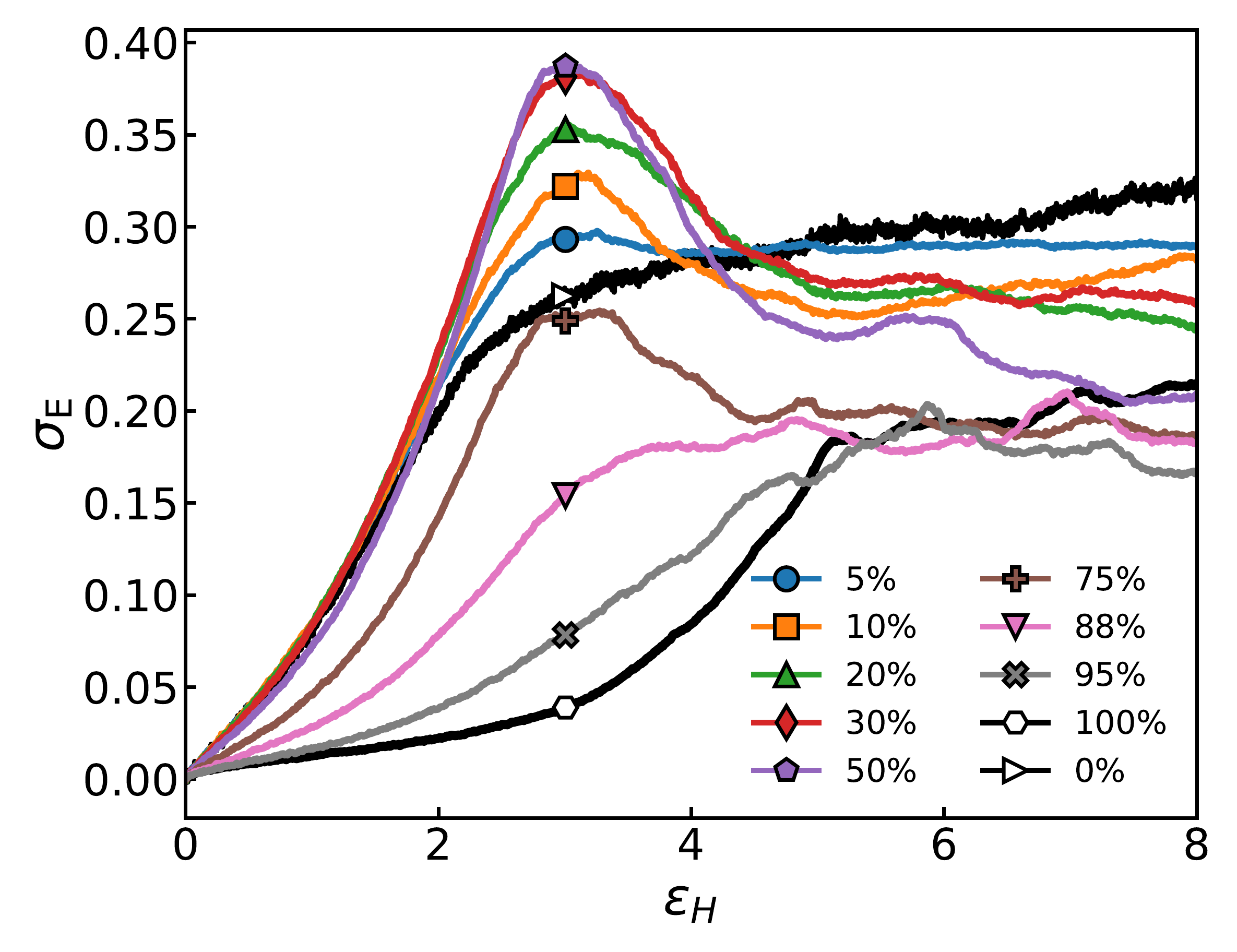}}
  \caption{
    Startup extensional stress growth curves for all blends and both pure melts elongated at the same Hencky strain rate.
    The strain rate $\dot{\varepsilon}_H=7.8\times 10^{-6}\ \tau^{-1}$ corresponds to a ring Weissenberg number $Wi^R=\tau_d^R\dot{\varepsilon}_H\approx1.1$ for $\phi_R=1.0$.
    For the neat linear chains, this corresponds to a \textit{Rouse} Weissenberg number of $\sim3$.
    Blends develop a prominent overshoot in $\sigma_E$ not observed in neat melts.
    The strain at the maximum stress corresponds closely to maximum extensibility $\lambda_{\mathrm{max}}=\sqrt{N_e/C_\infty}$ of the neat linear entanglement network.}
  \label{fig:overshoot}
\end{figure}

The startup extensional stress growth coefficient (transient stress) $\sigma_E = \sigma_{zz}-\sigma_{rr}$ versus the Hencky strain $\varepsilon_H$ for all eight blends (colors) as well as for the pure linear and pure ring melts (black) are plotted in Fig.~\ref{fig:overshoot}.
Here, $\sigma_{zz}$ and $\sigma_{rr}$ are the stress components parallel and perpendicular to the extension axis, respectively.
It is clear from Figure \ref{fig:overshoot} that the extensional stress in blends $\sigma_E(\phi_R,t)$ cannot be constructed by simply adding the stress evolution of pure melts of ring and linear chains.
Instead, consistent with Borger \textit{et al.},\cite{Borger2020} a prominent nonlinear overshoot develops.
The over shoot grows non-monotonically as $\phi_R$ increases from 0 and reaches a maximum peak stress at $\phi_R\approx 0.3$.
Notably we observe little change in the peak-stress in the range $\phi_R=0.3-0.5$, implying an unexpectedly broad range of blend compositions with a similar nonlinear startup.
Our analysis of chain conformations will show that this insensitivity of the peak stress to composition coincides with a range of $\phi_R$ where the stress contributions of elongated ring and linear chains are approximately equal. 
For $\phi_R>0.5$, the peak stress rapidly decreases towards the pure-ring response.
This corresponds to the $\phi_R$ where the mean squared displacement and PPA analysis observed a rapid change in ring diffusion and topological structure with  $\phi_R$, indicating the breakdown of the composite entanglement network.

\begin{figure}[h!]
  \centerline{\includegraphics[width=0.8\textwidth]{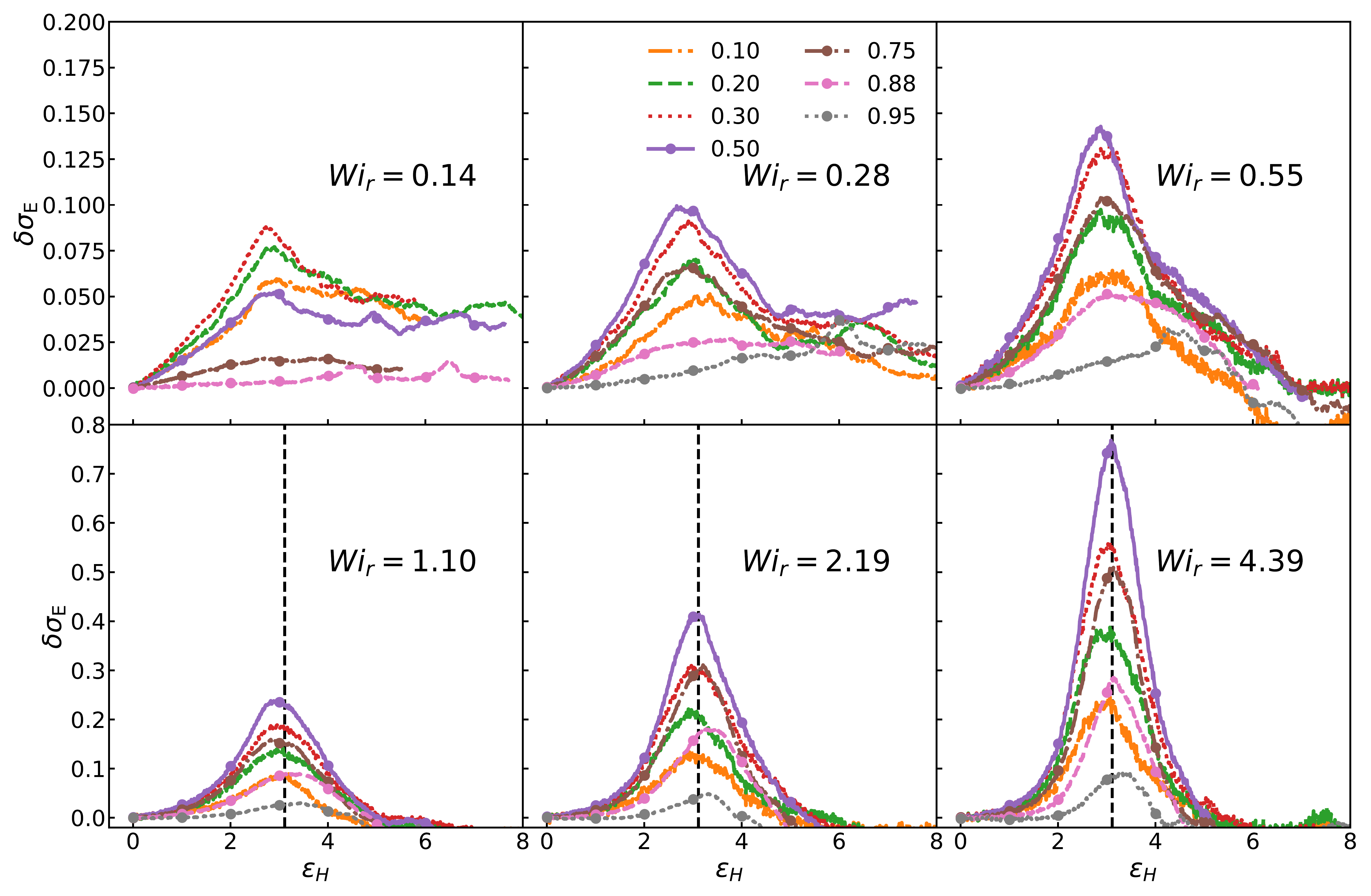}}
  \caption{
  Excess extensional stress $\delta\sigma_E$ in blends during startup uniaxial extensional flow at the indicated $Wi=\dot{\varepsilon}_H\tau_D^R(\phi_R=1)$.
  Excess stresses are computed relative to ideal mixing of startup data for neat melts at the same  strain rate.
  Data in (d) correspond to the $\sigma_E$ plotted in Figure \ref{fig:overshoot}.
  Vertical dashed lines indicate $\varepsilon_H=3.1$, the maximum extensibility of the neat linear entanglement network.
  }
  \label{fig:dsig}
\end{figure}

Blends with $\phi_R\geq0.88$ exhibit no apparent overshoot in stress as seen in Fig.~\ref{fig:overshoot}, but this is somewhat misleading because the extra stresses due to the coupling of ring and linear chains need to be considered relative to the response of their pure constituents. 
To do this, we can calculate a stress difference $\delta\sigma_E$ between the blend startup curves plotted in Fig.~\ref{fig:overshoot} and an ``ideal startup'' produced by simply mixing the startups for the neat melts (black curves):
\begin{equation}
    \delta\sigma_E(\phi_R,\varepsilon_H) = \sigma_E(\phi_R,\varepsilon_H) - \phi_R \sigma_E(1,\varepsilon_H) - (1-\phi_R)\sigma_E(0,\varepsilon_H)
    \label{eq:dsig}
\end{equation}
Figure \ref{fig:dsig}(d) plots $\delta\sigma_E$ curves for the data in Fig.~\ref{fig:overshoot} and for five additional strain rates in the other panels.
Shown this way, we observe an extra stress due to blending for all $\phi_R$ and strain rates.
As expected, the extra stress diminishes as $\phi_R$ increases --- leading to less ring-linear threading --- and as the strain rate increases --- inducing more elongation prior to unthreading.
The maximum of $\delta\sigma_E$ for most $\phi_R$ occurs at $\varepsilon_H\approx3.0$. 
This strain corresponds closely to the maximum extensibility of the linear entanglement network for this bead-spring model ($\varepsilon_H = \mathrm{ln}(\lambda_\mathrm{max})\approx3.1$).
At larger strains, entanglements relax nonaffinely and flow drives constraint release and unthreading, facilitating relaxation.
For $\phi_R\geq0.75$, there is a noticeable broadening of the peak in $\delta\sigma_E$ and the maximum shifts to larger strains for all strain rates. Both features are consistent with the decreasing density of topological constraints at large $\phi_R$.
This produces a sparser and less uniform network with a larger effective extensibility, broadening the peak. 

\subsection{Chain Statistics During Uniaxial Elongation}

Molecular simulations enable us to relate the extensional stress growth in Fig.~\ref{fig:overshoot} to changes in the chain conformations of ring and linear polymers. Models of polymer dynamics typically represent chain conformations in terms of the average orientation and stretch of chain segments relative to equilibrium.\cite{Dealy2018} Both elongation and orientation order parameters can be derived from the internal distance vector $\vec{R}(n)$ which gives the distance between monomers separated by $n$ bonds. The average elongation is described by the extension ratio of a segment $h(n)=\langle R(n)\rangle/nb$ where $b$ is the average bond length. As segments approach full extension, $h(n)\rightarrow1$. For extensional flows where flow orients conformations along the extension axis, it is convenient to measure segment orientation with the nematic orientation order parameter $S(n)=1.5(\langle \mathrm{cos}\ \theta(n)\rangle-0.5)$, where $\theta(n)$ is the angle of $\vec{R}(n)$ relative to the extension axis. $S(n)=0$ for isotropically oriented segments and $S(n)=1$ for fully oriented segments.
Since most models assume the macroscopic stresses in both linear and ring polymer melts can be captured by the statistics of entanglement-scale segments, we use $n=N_e$ in what follows. 
Studying entanglement-scale segments is also convenient because ring and linear chains can be treated similarly.

\begin{figure}[h!]
  \centerline{\includegraphics[width=0.5\textwidth]{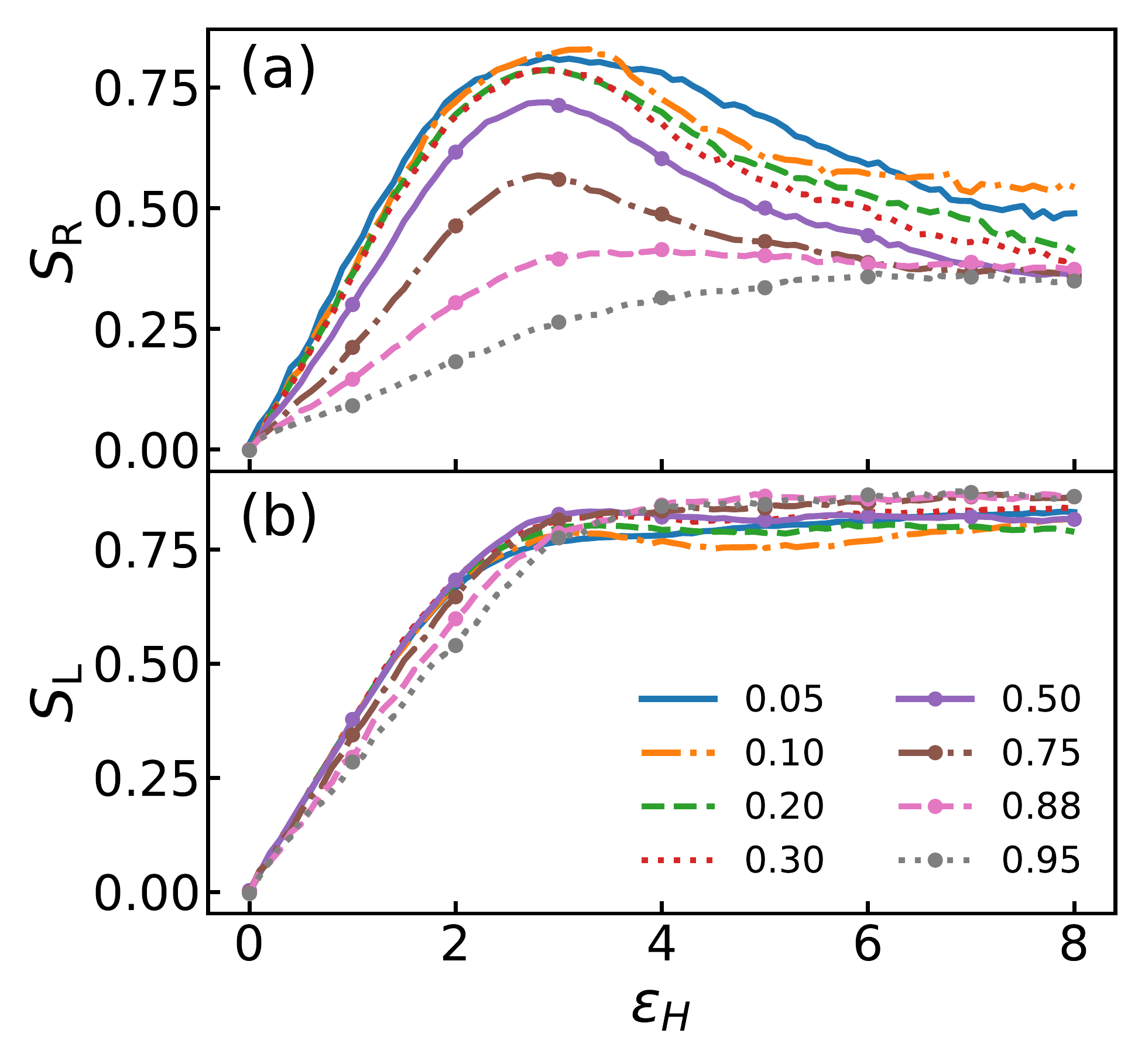}}
  \caption{Average nematic orientational order of ring (a) and linear (b) entanglement segments with length $N_e$ during startup uniaxial elongation flow for all blends. Linear chain orientation shows little change in startup evolution with varying $\phi_R$, while rings develop a prominent overshoot as $\phi_R$ decreases.}
  \label{fig:orientation}
\end{figure}

In Figure \ref{fig:orientation}(a)-(b), the nematic orientation of entanglement segments during extensional flows is plotted for ring ($S_r$) and linear ($S_\ell$) chains, respectively.
It is immediately obvious that the two architectures have distinct orientational dynamics during elongation.
Linear chains in blends show a monotonic growth of orientation with increasing strain and almost no sensitivity to $\phi_R$.
Linear chain orientation saturates at $S_L \sim 0.8$, which is essentially the same response as is seen for pure linear melts at the same $Wi_R$ \cite{OConnor2018PRL}.
This is consistent with our  analysis of the terminal time $\tau_d^L$ for linear chains in Figure \ref{fig:eqlstats}(c).
Linear chains relax orientation through reptation on timescales $\sim\tau_d^L$, both in equilibrium and highly elongated states.\cite{OConnor2019} 
Thus, the  insensitivity of $S_L$ to $\phi_R$ can be understood by the similar insensitivity of $\tau_d^L$. 
Evidently, linear chains reptate similarly through ring and linear matrices both in and out of equilibrium.

Unlike linear chains, the orientation  $S_R$ of the rings in elongated blends show a strong dependence on $\phi_R$ and exhibit an overshoot in orientation with a peak emerging at $\varepsilon_H\approx3.0$, which coincides with the stress overshoot in Figure \ref{fig:overshoot}.
The peak in entanglement orientation grows as $\phi_R$ decreases below 0.88 and the peak of the overshoot saturates at a maximum orientation $S_R\sim0.8$ once $\phi_R<0.5$.
This maximum value of $S_R$ is similar to the steady orientation of the linear chains.
In addition, as $\phi_R$ decreases below 0.5, the increase in $S_R$ at small strains rapidly approaches the same curve followed by $S_L$ for linear entanglements.
Both of these observations suggest that as rings become highly-threaded into the linear entanglement network ($\phi_R\rightarrow 0$), the ring entanglement segments behave like linear entanglement segments and deform similarly within the entanglement network.
Indeed, comparing Figures \ref{fig:orientation} and \ref{fig:eqlstats}(b), we see that the saturation in ring orientability coincides with the slow-down in growth of the primitive path length $L_{pp}^R$ with decreasing $\phi_R$.
The rings cannot maintain this peak orientation because the elongation flow also drives the convective release of ring-linear threadings. As described by Borger et al. \cite{Borger2020}, the rapid removal of these threadings decouples rings from the entanglement network, allowing rings to recoil and substantially relax their orientation much faster than their equilibrium relaxation time.
A similar ring recoil due to convective release of linear threadings has also been observed in single-molecule experiments of lightly threaded ring-linear solutions.\cite{Zhou2019,Young2020,Zhou2021}

\begin{figure}[h!]
  \centerline{\includegraphics[width=0.5\textwidth]{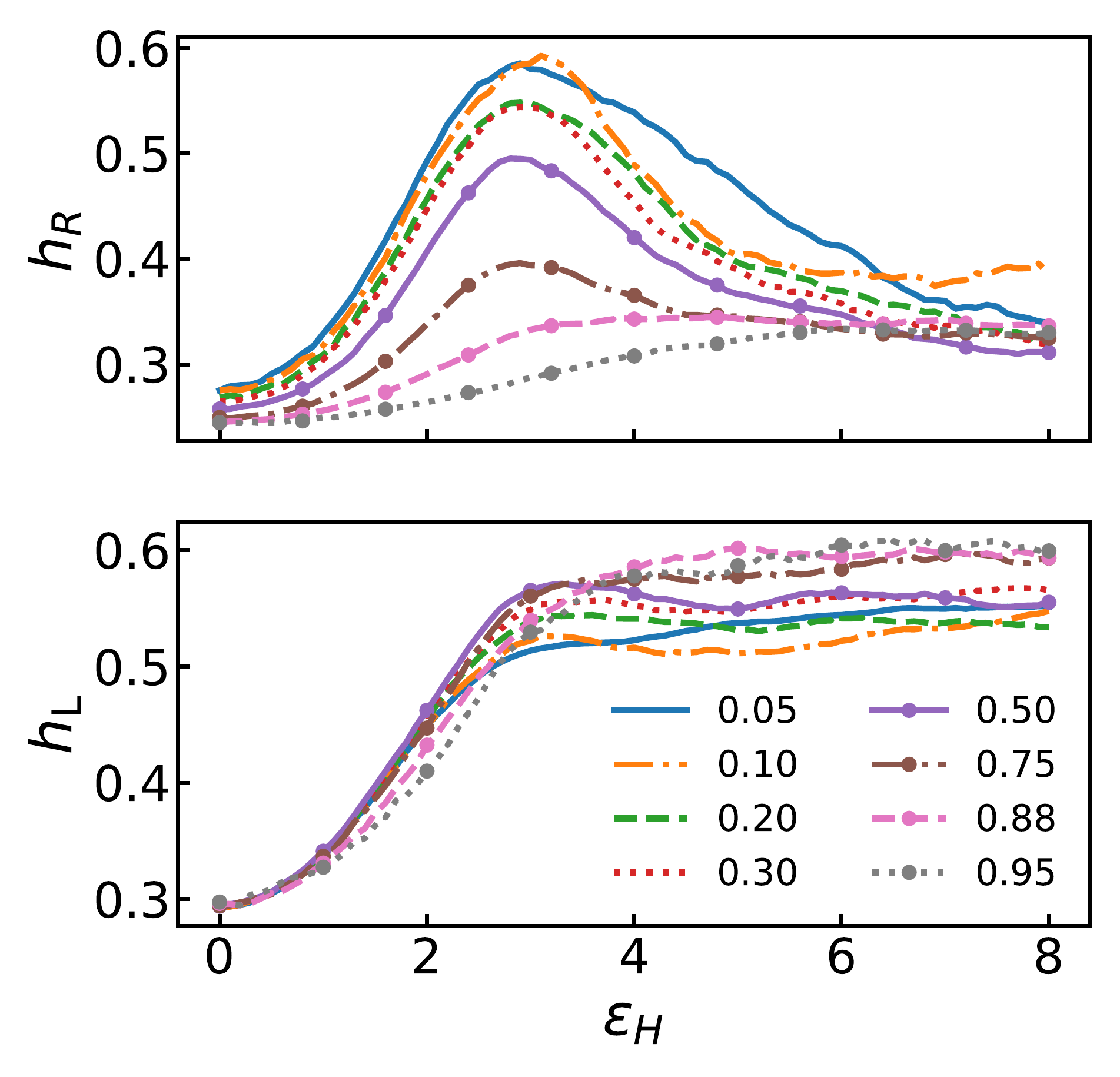}}
  \caption{Average extension ratios for (a) ring and (b) linear entanglement segments with length $N_e$ during startup uniaxial elongation flow for all blends. Similar to $S_R$ and $S_L$, linear segments show relatively little change in startup evolution with varying $\phi_R$, while rings develop a prominent overshoot as $\phi_R$ decreases.}
  \label{fig:stretch}
\end{figure}

The evolution of segment stretch is qualitatively similar to segment orientation. Figure \ref{fig:stretch} plots linear $h_L$ and ring $h_R$ extension ratios versus strain for all blends. 
The elongation of linear entanglements again shows a weak dependence on $\phi_R$. For all blends, $h_L$ increases monotonically to a steady value of  $\sim0.55$, which is slightly higher ($\sim0.6$) for $\phi_R\geq0.75$.
For rings, an overshoot again develops as $\phi_R$ decreases and its peak saturates once $\phi_R<0.5$ to an extension ratio $h_R\sim0.55$ similar to the steady extension ratio of linear segments.

We note that while $S_L$ and $h_L$ show a relatively weak dependence on $\phi_R$, they do exhibit some subtle and interesting features. For the largest $\phi_R$ near 1.0, both $S_L$ and $h_L$ show a slower increase with $\varepsilon_H$ but also achieve their largest steady values.
The weaker increase with strain is expected due to the deterioration of the entanglement network at large $\phi_R$. However, the larger steady values are somewhat surprising since Figure \ref{fig:eqlstats} shows that the linear chain relaxation time \textit{decreases} with increasing $\phi_R$.
This behavior appears to be another consequence of ring-linear threading during nonlinear flow. 
At large $\phi_R$, each linear chain threads many rings, as can be seen in Figure \ref{fig:networks} for $\phi_R=0.95$. 
These threadings do not significantly slow the diffusion of the linear chains in equilibrium.
However, when threaded rings are convected by the flow, they can induce large deformation of the linear chain threading them as it is dragged out.
Single-molecule experiments have also observed threading-induced overstretching of linear chains in semidilute blends.\cite{Young2020,Zhou2021}

\begin{figure}[h!]
  \centerline{\includegraphics[width=0.5\textwidth]{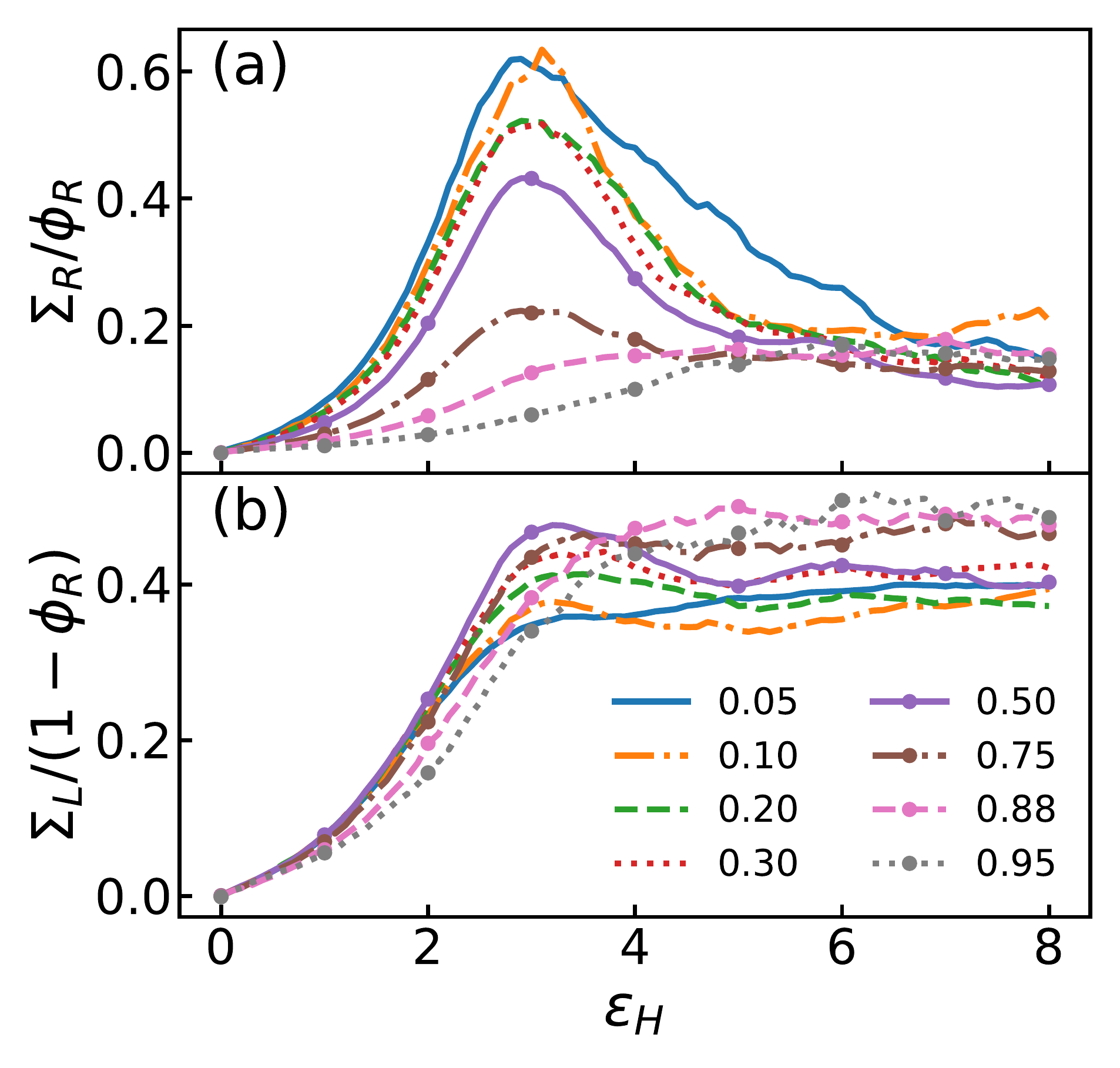}}
  \caption{Normalized entropic stress contributions computed from Eq. \ref{eq:ent} for entanglement segments of ring (a) and linear (b) chains during startup uniaxial elongation flow for all blends. $\Sigma_R$ and $\Sigma_L$ are normalized by the volume fraction of their respective species so that the curves give the stress contributions per entanglement segment. }
  \label{fig:ent}
\end{figure}

The above analysis shows a direct link between the $\phi_R$ dependence of the extensional stress overshoot (Figure \ref{fig:overshoot}) and the nonlinear dynamics of the threaded ring polymers.
We make this connection more explicit by computing the approximate entropic stress contributions of linear and ring entanglement segments from their respective chain statistics.
Following recent work \cite{OConnor2018PRL,OConnor2019,OConnor2020}, we approximate the extensional component of the entropic stress of segments as
\begin{equation}
    \Sigma = \phi\frac{\rho k_B T}{n_K}\langle h L^{-1}(h) S\rangle
    \label{eq:ent}
\end{equation}
where $n_K\approx1.88$ is the number of monomers in a Kuhn segment and $\phi$ represents either $\phi_R$ for ring segments or $1-\phi_R$ for linear segments.
Figure \ref{fig:ent} (a) and (b) plots normalized entropic stress contributions $\Sigma_R/\phi_R$ and $\Sigma_L\/(1-\phi_R)$ for ring and linear entanglement segments, respectively.
Normalized by the volume fraction, this quantity approximates the per-segment contribution to the entropic stress.
The dependence of $\Sigma_R$ and $\Sigma_L$ on strain and composition track closely with the conformational statistics, confirming that rings dominate the behavior of the extensional stress overshoot.

Figure \ref{fig:ent} also reveals why there is little change in the peak value of $\sigma_E$ between $\phi_R=0.5$ and $0.3$ (Figure \ref{fig:overshoot}).
Increasing $\phi_R$ replaces linear segments with ring segments, and we expect the change in $\sigma_E$ with $\phi_R$ to be $\sim\Sigma_R/\phi_R - \Sigma_\ell/(1-\phi_R)$.
Simply put, $\sigma_E$ should show little change with $\phi_R$ when ring and linear segments contribute similar amounts.
Figure \ref{fig:ent}(b) shows that linear segments at all $\phi_R$ give a per-segment stress contribution of $\sim0.4$ at the stress overshoot ($\varepsilon_H=3.0)$.
The stress contribution of rings at $\varepsilon_H=3.0$ vary with $\phi_R$. At low $\phi_R$, rings contribute more to the overshoot than linear segments.
Ring contributions decrease with increasing $\phi_R$ such that $\Sigma_R/\phi_R$ is $\sim\Sigma_L/(1-\phi_R)$ in the region $\phi_R=0.3-0.5$ where there is little change in peak stress.
Notably, this range of compositions includes the $\phi_R$ where both Halverson \textit{et al.} (in simulations) and Roovers (in experiments) report maximums in the zero-shear viscosity of their ring-linear blends with similar $Z$.\cite{halverson12,Roovers1988}
It would be interesting to investigate whether a correspondence can be drawn between the peak in $\sigma_E$ during nonlinear extensional flow and the peak in $\eta_0$ measured by SAOS for entangled ring-linear blends.

\begin{figure}[h!]
  \centerline{\includegraphics[width=0.75\textwidth]{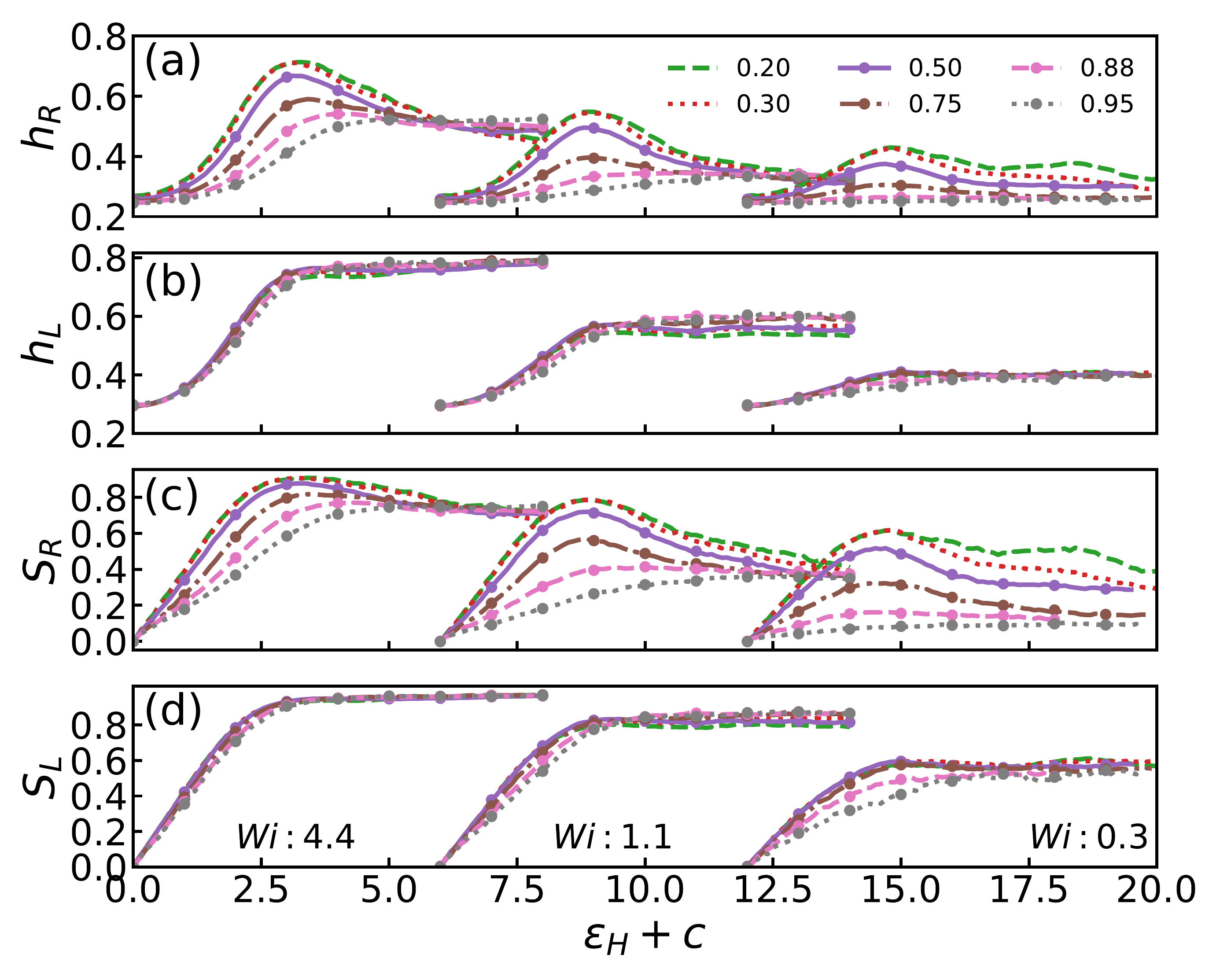}}
  \caption{Aggregated stretch (a),(b) and orientation (c),(d) statistics for ring and linear chains, respectively, during startup uniaxial elongation for six blends and three strain rates.
  The $\phi_R$ for the six blends are indicated in (a) and data are shifted horizontally so that (a)-(d) form columns corresponding to the different Weissenberg numbers $Wi=\dot{\varepsilon}_H \tau_D^R(\phi_R=1)$ in panel (d).
  The center column is the same data plotted in Fig.~\ref{fig:orientation} and \ref{fig:stretch}.
  }
  \label{fig:statrate}
\end{figure}

Figure \ref{fig:statrate} summarizes stretch and orientation data for three different strain rates and six values of $\phi_R$.
Curves in each panel are offset by strain increments of six, such that they form columns corresponding to different strain rates.
The central data correspond to the results plotted in Figures \ref{fig:orientation} and \ref{fig:stretch}. 
The data to the left and right correspond to $\dot\varepsilon_h$ four times large and smaller, respectively.
This captures results for Weissenberg numbers both above and below 1 for both linear and ring populations.

The results at different rates are qualitatively similar. 
In all cases, linear chain statistics increase monotonically and plateau to a steady-state stretch ratio or orientation that increases towards 1.0  with increasing rate. 
At all rates, linear chain statistics show relatively little sensitivity to $\phi_R$, except for a slight change in shape for the largest $\phi_R\geq0.88$, as previously discussed.
The ring stretch ratio and orientation develop an overshoot as $\phi_R$ decreases for all three rates. 
The peak of the overshoot increases with rate and saturates for $\phi_R<0.5$ at all rates.
The qualitatively similarities of the overshoot at different rates are consistent with the work of Borger \textit{et al.} \cite{Borger2020}.
Their experiments and simulations for a $\phi_R=0.3$ blend found a qualitatively similar overshoot behavior for a wide range of strain rates.
This seems reasonable since the overshoot is driven by threaded rings which we have shown have a relaxation time that is significantly larger than for linear chains (Figure \ref{fig:eqlstats}(c)).
Thus, most practical flow rates will always be strongly nonlinear for the threaded rings and should produce a qualitatively similar overshoot in stretch and orientation.

\begin{figure}[h!]
  \centerline{\includegraphics[width=0.5\textwidth]{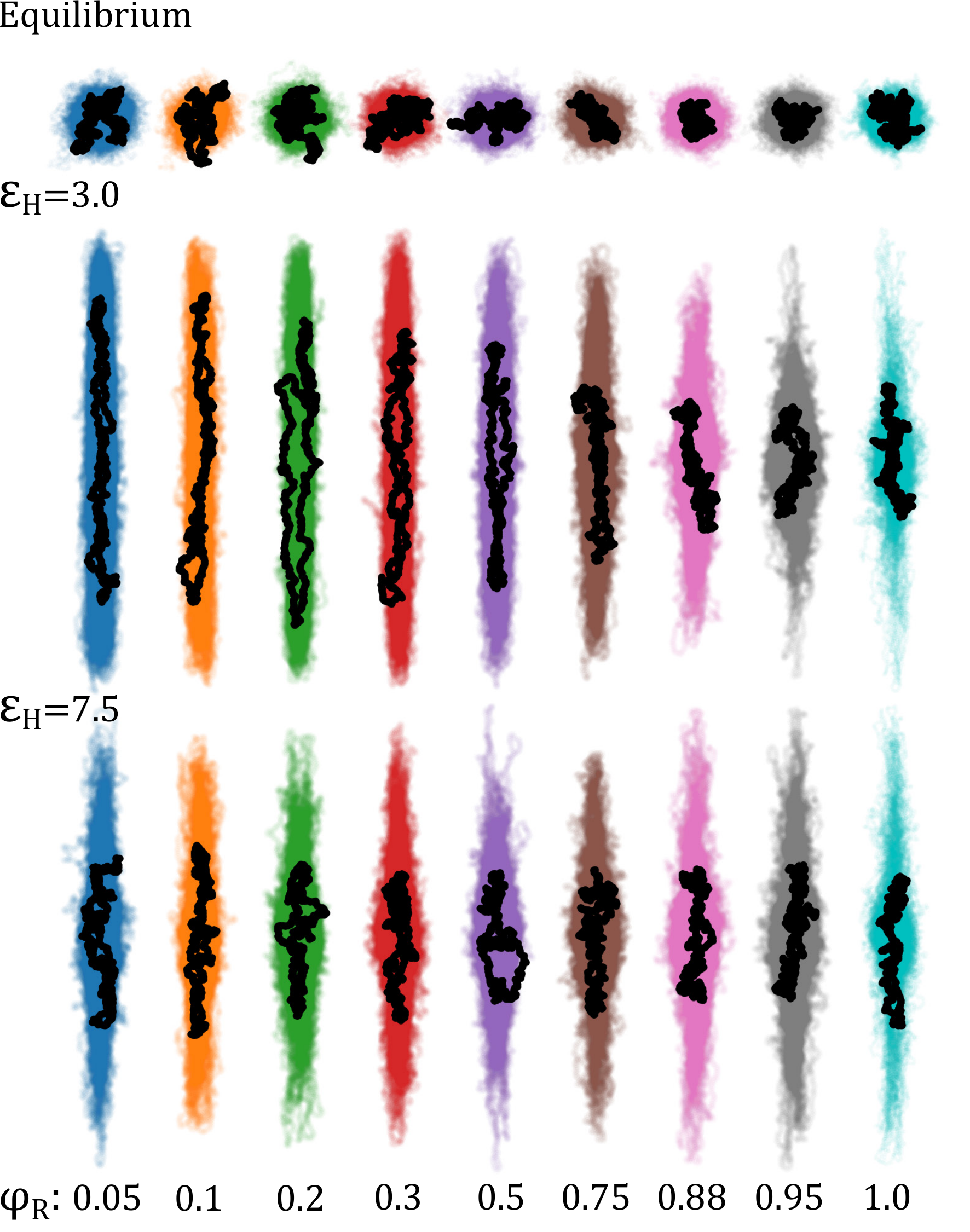}}
  \caption{Ensembles of ring polymer conformations for all blends in equilibrium, at the peak of the stress overshoot ($\varepsilon_H=3.0$), and after recoil ($\varepsilon_H=7.5$). 
  Similar to Fig.~\ref{fig:eqlconf}, ensembles are generated by plotting translucent chains shifted to a common center of mass, producing a colored cloud. 
  The black chain emphasized in each ensemble is the ring with $R_g$ or closest to the root-mean-square value at that strain.
  Rings exhibit substantial recoil between $\varepsilon_H=3.0$ and 7.5 that diminishes as $\phi_R$ increases.}
  \label{fig:ringconf}
\end{figure}

The average statistics of rings $h_R$, and $S_R$ clearly capture the overshoot in $\sigma_E$ and its dependence on $\phi_R$.
However, to understand the molecular mechanisms driving the overshoot, it is informative to visualize the full ensemble of ring polymers during elongational flow.
Similar to Figure \ref{fig:eqlconf}, Figure \ref{fig:ringconf} shows the ensembles of ring polymer conformations for all $\phi_R$ at three values of $\varepsilon_H$ during the $Wi_r\approx1.1$ startup flow. As before, the colored clouds are the superimposed ensembles of ring conformations shifted to a common center of mass and the black chain in each ensemble is the ring with a span closest to the ensemble average.
Ensembles are shown in equilibrium, at the overshoot strain $\varepsilon_H=3.0$, and well after recoil at $\varepsilon_H=7.5$. 

\begin{figure}[h!]
  \centerline{\includegraphics[width=0.5\textwidth]{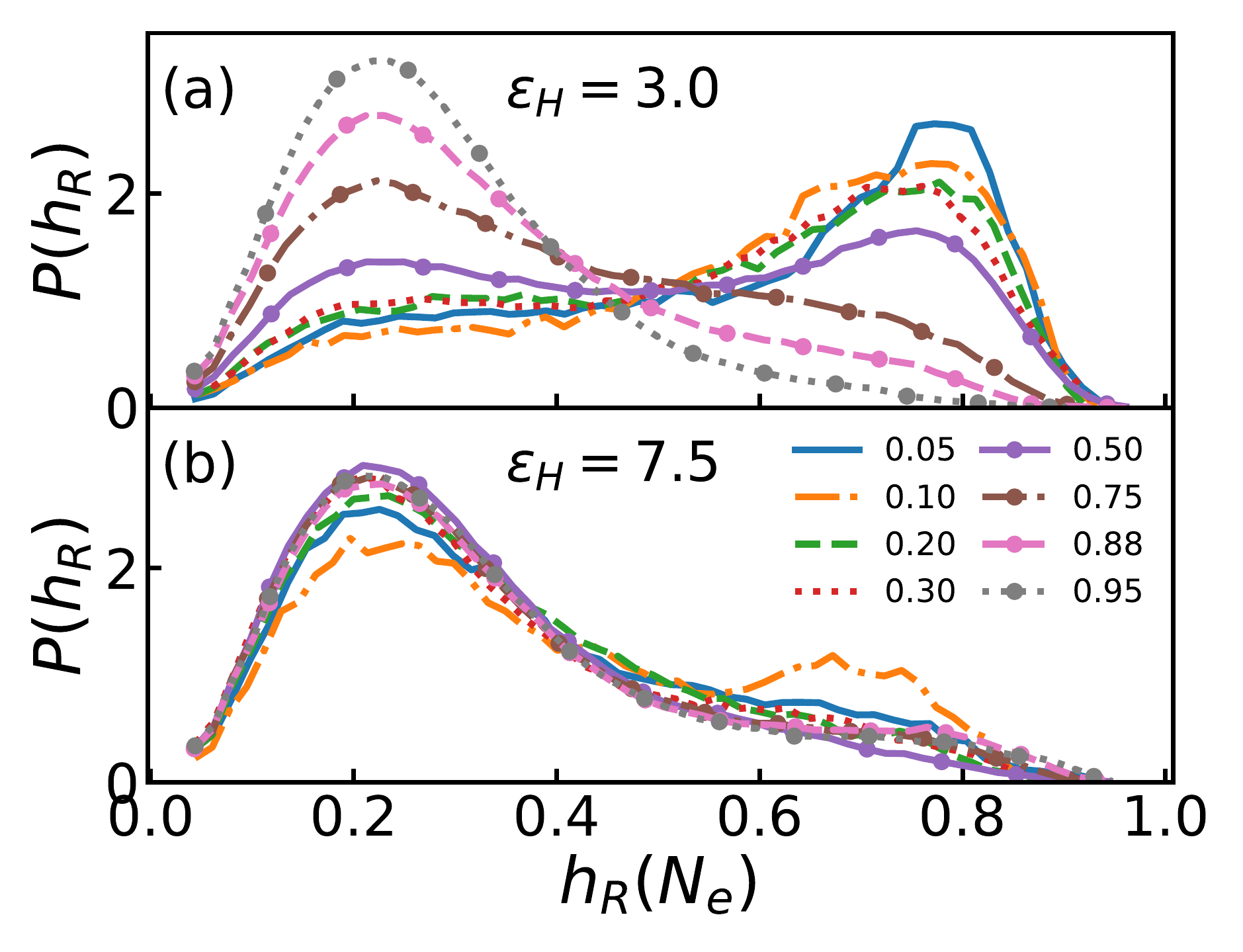}}
  \caption{Probability distributions of the extension ratio of ring entanglement segments $h_R=R(N_e)/N_e b$ at (a) the overshoot strain $\varepsilon_H=3.0$ and (b) after recoil at $\varepsilon_H=7.5$.}
  \label{fig:hhist}
\end{figure}

Comparing ensembles in equilibrium and at $\varepsilon_H=3.0$, we can see a qualitative change in the distribution of elongated as $\phi_R$ increases.
This change is quantified by entanglement stretch distribution functions $P(h_R)$, shown for $\varepsilon_H=3.0$ and $7.5$ in Figure \ref{fig:hhist}(a) and (b), respectively.
For blends with $\phi_R \leq 0.5$, the elongated ensembles at $\varepsilon_H=3.0$ are similar, corresponding to the saturation in the peak stretch in Figure \ref{fig:stretch}. 
At these compositions, rings deform with the linear entanglement network up to its maximum extensibility $\varepsilon_H\sim3.1$.
This behavior is reflected in the prominent peak in $P(h_R)$ at $h_R\sim0.8$ in Fig.~\ref{fig:hhist}(a).
As $\phi_R$ increases above 0.5, the entanglement network deteriorates and the elongated ensembles become noticeably skewed, with most rings staying near their equilibrium conformations and a smaller number of rings becoming highly elongated at $\varepsilon_H=3$.
This is reflected in the $P(h_R)$ for $\phi_R>0.5$ by a prominent peak near the equilibrium value of $h_R$ with a long shoulder extending toward large values.

The dramatic differences in response of the stretch distributions at strains below the overshoot corresponds to different mechanisms mediating the elongation of rings. 
At low $\phi_R$, rings are highly threaded by linear chains (Figure \ref{fig:eqlstats}(d)) and are well-embedded in the linear entanglement network such that $\tau_R > \tau_\ell$ (Figure \ref{fig:eqlstats}).
Because of this, rings deform affinely with the linear entanglement network until it is fully extended and rings unthread by convective constraint release.
As $\phi_R$ increases above 0.5, the linear entanglement network deteriorates and can no longer mediate the elongation of rings. 
Instead, rings begin to behave like pure ring melts, and develop highly skewed stretch distributions. 
These distributions are caused by rings linking together through self-threadings to form long, supramolecular daisy chains, as previously described by O'Connor \textit{et al.}.\cite{OConnor2020}

Figure \ref{fig:ringconf}(c) shows ring ensembles at $\varepsilon_H=7.5$ after flow has driven ring-linear unthreading. 
Here, blends with prominent stress overshoots ($\phi_R\leq0.5$), exhibit significant reconfiguration as rings unthread and recoil.
Considering the $P(h_R)$ distributions in Figure \ref{fig:hhist}(b), we observe that the recoil corresponds to the disappearance of the peak at large $h_R$ and the appearance of a peak near the equilibrium value of $h_R$.
In other words, the highly threaded rings that approach full extension completely recoil back to their equilibrium sizes after unthreading.
For $\phi_R\geq0.75$, unthreading also occurs but its effect is less pronounced due to the combination of rings elongating through ring-linear threadings and ring-ring threadings.
We observe the high-$h_R$ shoulder for $\phi_R=0.75$ and 0.88 diminish with a corresponding increase in their equilibrium peaks.
Only the $\phi_R=0.95$ system shows no significant recoil. 
Its equilibrium peak diminishes between $\varepsilon_H=3.0$ and $7.5$, while its should increases and extends to larger $h_R$.
This is due to the slow increase in  rings self-threading to form supramolecules. 

Notably, $P(h_R)$ for all blends recoil to a similar distribution by $\varepsilon_H=7.5$.
While interesting, it is clear from the transient stress and chain statistics in Figures \ref{fig:overshoot}, \ref{fig:orientation}, and \ref{fig:stretch} that the blends have not achieved steady-states or steady statistical distributions by $\varepsilon_H=8.0$.
Achieving a steady state would require a balance of threading formation and release, which is likely a slow process.
Indeed for pure rings, the spontaneous formation of supramolecular chains by ring-ring self-threading during flow implies that there may not be a true steady-state at all.
Thus this correspondence at $\varepsilon_H=7.5$ is likely transient and unlikely to be achievable in experiments, regardless.
Nonetheless this correspondence does raise interesting questions about the transient relaxation of blends towards equilibrium after unthreading drives large changes in topological structure.
It would be interesting to learn if  elongated blends exhibits domains of convergent relaxation, as observed in highly elongated linear melts.\cite{OConnor2019} We will pursue these questions in future works.

\subsection{Blend Topology During Elongation}
Throughout this paper we have rationalized the extensional rheology and dynamics of ring-linear blends in terms of ring-linear threadings and their unthreading during flow.
The idea of linear chain ``threadings'' has a long history\cite{Mills1987,Roovers1988,McKenna1989} and has grown more common in the recent literature.\cite{Zhou2019,Smrek2019,Smrek2020,Michieletto2021}
Regardless, in a molecular simulation study like this one, there is no reason for us to require a reader to take this view for granted.
Not when we can characterize it directly.
So, to wrap up, we have used primitive path analysis to characterize the evolution of the topological structure of the ring-linear entanglement network during uniaxial elongation.\cite{Everaers2004,Halverson2012}
This allows us to both visualize the ring unthreading transition identified by Borger et al. \cite{Borger2020}, and quantitatively explore the changes in topological structure associated with it.

\begin{figure}[h!]
  \centerline{\includegraphics[width=0.75\textwidth]{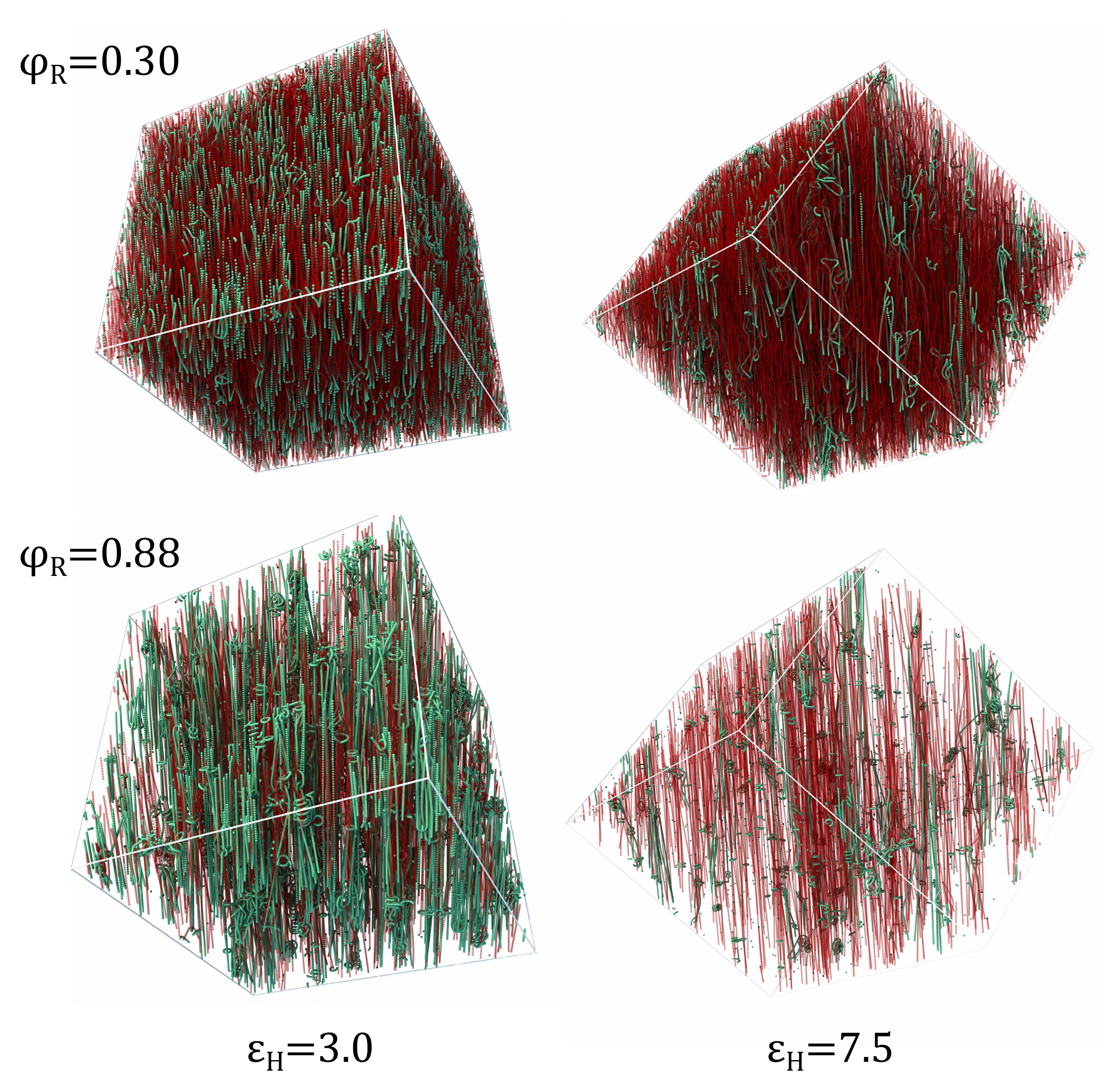}}
  \caption{  Visualization of ring unthreading during extensional flow with blend network topologies generated by primitive path analysis for two ring-linear blends with $\phi_R=0.3$ (top) and 0.88 (bottom).
  Ring and linear primitive paths are colored green and red, respectively.
  Configurations are shown at the overshoot strain $\varepsilon_H=3.0$ where ring primitive paths are elongated along the extension axis (left), and after the rings have recoiled at $\varepsilon_H=7.5$ (right).
  The flow-driven unthreading of the rings produces significant and visible changes in the network structure for both $\phi_R$.
  }
  \label{fig:unthread}
\end{figure}

Figure \ref{fig:unthread} shows primitive-path network configurations similar to Figure \ref{fig:networks} for a $\phi_R=0.30$ and $0.88$ blend at $\varepsilon_H=3.0$ and $7.5$, before and after the overshoot.
As before, these configurations are generated by removing intramolecular steric repulsion from all polymers while fixing the chain ends of just the \textit{linear} chains in space.
Ring polymers are not constrained and will thus collapse into points unless they are held open by threading constraints.
Even though one blend exhibits a prominent peak in extensional stress (Figure \ref{fig:overshoot}) and the other does not, it is clear from the figure that both possess a significant number of threaded rings (green) and that their primitive path statistics undergo significant changes between $\varepsilon_H=3.0$ and $7.5$.
At $\varepsilon_H=3.0$, many ring primitive paths in both blends are extended along the extension axis.
At $\varepsilon_H=7.5$, the majority of ring primitive paths have recoiled dramatically, decreasing in both length and orientation.
This is a direct view of the topological changes produced by the the convective unthreading of the blends during elongation.

For both blends, some ring paths remain elongated even at $\varepsilon_H=7.5$, but there is a notable difference in their topological constraints.
For the $\phi_R=0.3$ blend, each ring has $N_t\approx15$ linear threadings in equilibrium, and the ring paths that remain extended after the overhsoot are held open by linear threadings that have survived out to $\varepsilon_H=7.5$.
In contrast, the $\phi_R=0.88$ blend has very few linear threadings in equilibrium $N_t<1$, and we observe that most extended ring paths that surive to $\varepsilon_H=7.5$ are also self-threaded with other rings to form supramolecular daisy chains \cite{OConnor2020}. 
These supramolecular objects extend strongly in the flow, which evidently allows them to more effectively entangle with and be threaded by linear chains.

\begin{figure}[h!]
  \centerline{\includegraphics[width=0.5\textwidth]{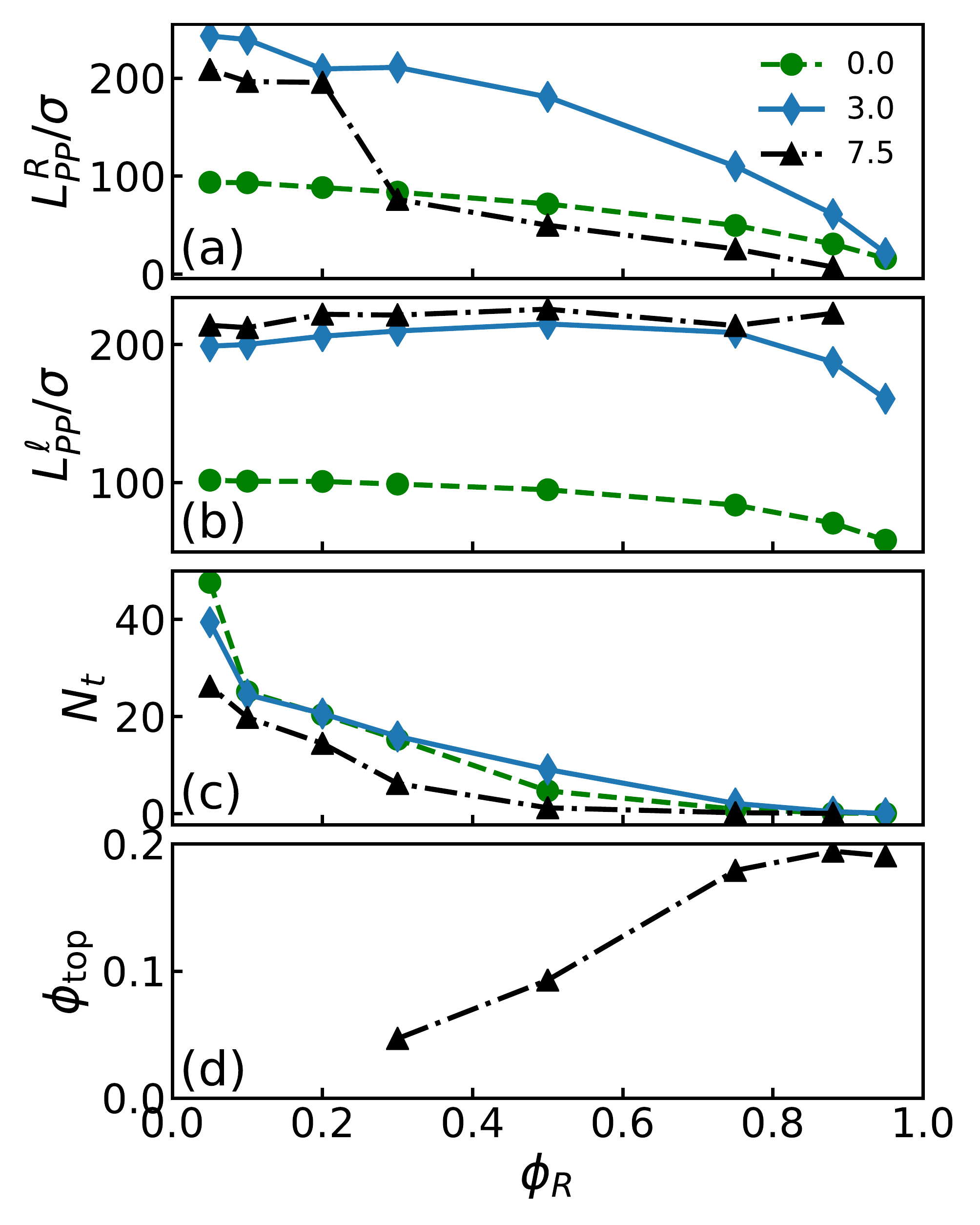}}
  \caption{
  Topological chain statistics derived from primitive path analysis of ring-linear blends in equilibrium and during uniaxial elongation flow.
  Average primitive path lengths for (a) ring and (b) linear chains for all $\phi_R$ at $\varepsilon_H=0$, 3.0, and 7.5.
  (c) the average number of linear chains threading a ring $N_t$ at the same three strains.
  (d) The fraction $\phi_{\rm top}$ of rings participating in ring daisy chains via ring self-threadings at $\varepsilon_H=7.5$ only. Self-threading data for $\phi_R<0.3$ is not computed due to limited data.
  }
  \label{fig:flowtop}
\end{figure}

Figure \ref{fig:flowtop} quantifies the evolving topological statistics as a function of $\phi_R$ during elongation. Panels (a) and (b) plot the root mean squared primitive path lengths $L_{pp}^{R}$ and $L_{pp}^L$ of ring and linear polymers, respectively, in equilibrium and at $\varepsilon_H=3.0$ and $7.5$.
Panel (c) plots the average number of linear threadings per ring $N_t$ for all $\phi_R$ at the same three strains.
Considering $L_{pp}^L$ in panel (b), we see that linear primitive path lengths increase monotonically with strain at all $\phi_R$.
They increase substantially between $\varepsilon_H=0.0$ and 3.0, and slightly more between 3.0 and 7.5.
This expected, since $L_{pp}^L$ tracks with the elongation of linear conformations, which we have already shown increase monotonically for all $\phi_R$.

The changes for rings are more interesting. $L_{pp}^R$ increases substantially between equilibrium and $\varepsilon_H=3.0$ and then decreases between $\varepsilon_H=3.0$ and 7.5.
For $\phi_R$ near zero, the magnitude of the increase in $L_{pp}^R$ at the overshoot is similar to that for linear chains, but is slightly larger. 
This is consistent with slower relaxation dynamics ($\tau_D^R>\tau_D^L$) of well embedded rings relative to linear chains (Figure \ref{fig:eqlstats}).
$L_{pp}^R$ at $\varepsilon_H=3.0$ decreases as $\phi_R$ increases towards 1.0.
This is essential since, as we've seen, the equilibrium threadings per ring $N_t$ is $\sim1$ or less for $\phi_R>0.5$, and thus few rings have enough linear threadings to couple $L_{pp}^R$ to the affine flow field.

$L_{pp}^R$ decreases between $\varepsilon_H=3.0$ and 7.5 for all $\phi_R$, but the size of this drop varies significantly.
In particular, for $\phi_R\leq0.20$, $L_{pp}^R$ drops by only $\sim20\%$ by $\varepsilon_H=7.5$.
Meanwhile, blends with larger $\phi_R$ show a much larger drop in $L_{pp}^R$, with it dropping below its equilibrium value in all cases.
This large drop at larger $\phi_R$ is consistent with the significant visible change in the $\phi_R=0.88$ network in Figure \ref{fig:unthread}.

The recoil of $L_{pp}^R$ is driven by the convective release of ring-linear threadings. This is shown in the plots of $N_t$ in Figure \ref{fig:flowtop}(c), and these curves offer some insight about the variation in the observed recoil. 
For most $\phi_R$, there is relatively little change in $N_t$ between equilibrium and $\varepsilon_H=3.0$, but all exhibit a significant drop in $N_t$ after the overshoot to values below equilibrium.
What appears to distinguish the blends with small and large recoil in $L_{pp}^R$ is the number of threadings that survive after recoil at $\varepsilon_H=7.5$.
For $\phi_R\leq0.2$, rings have on average $N_t>19$ threading constraints at $\varepsilon_H=7.5$. This is greater than the $Z\sim18$ binary entanglements formed by one of the linear chains within a linear entanglement network.
Thus, even after substantial unthreading, rings in these systems remain highly constrained.
In contrast, $N_t$ at $\varepsilon_H=7.5$ drops rapidly for $\phi_R>0.2$ and is $\sim1$ or less for $\phi_R>0.5$. 
This represents an almost complete unthreading of rings, such that many ring primitive paths will collapse into points, driving $L_{pp}^R$ below its equilibrium value.

Figure \ref{fig:flowtop}(a)-(c) show that ring primitive paths collapse and decouple from the background flow as $\phi_R\rightarrow1$ and $N_t\rightarrow0$.
However, we know that even for $\phi_R=1.0$, rings still elongate and produce significant stresses in extensional flows (Figure \ref{fig:overshoot}) due to their linking into supramolecular daisy-chains that do not directly contribute to the measure of $L_{pp}^R$.
Indeed, in O'Connor et al. \cite{OConnor2020}, we studied the same pure ring melt considered here, and observed that about $20\%$ of the rings self-threaded into supramolecular clusters at this strain rate ($Wi\approx1.1$).
As in that work, we can measure the fraction of rings self-threaded into such structures by applying a slightly modified PPA.
In this analysis, we constrain the bead on each ring that is closest to that ring's center of mass.
We then remove remove all intramolecular steric interactions for rings and \textit{delete} all linear chains from the configuration.
This protocol drives all rings to collapse into their constrained bead, such that only rings linked to other rings through self-threadings will remain open.
After this procedure, we compute the fraction $\phi_{\mathrm{top}}$ of rings with a gyration radius $R_g > 1.0\sigma$. $\phi_{\mathrm{top}}$ sets an upper bound on the fraction of rings participating in supramolecules.

Calculations of $\phi_\mathrm{top}$ at $\varepsilon_H=7.5$ for $\phi_R\geq0.3$ are plotted in Figure \ref{fig:flowtop}(d).
For lower $\phi_R$, rings are too dilute to produce good self-threading statistics for our system sizes and simulation durations.
We observe a steady increase from $\phi_{\mathrm{top}}\sim0.05$ at $\phi_R=0.3$ to $\phi_{\mathrm{top}}\sim0.18$ at $0.75$. The increase then slows as it comes near to  $\phi_{\mathrm{top}}\sim0.20$ as previously reported for $\phi_R=1.0$.\cite{OConnor2020}
In order to linkup, rings must overlap and diffusively self-thread through each other.
Thus, the steady growth in ring linking is likely due to a combination of increased ring-ring overlap and increased ring diffusion as linear threading constraints are removed.
The accelerated diffusion is quantified by $\tau_d$ in Figure \ref{fig:eqlstats}(c). 
The increasing overlap of rings is subtle since, as shown in Figure \ref{fig:eqlstats}(a), the size of rings decreases significantly as $\phi_R$ increases and threadings are removed. 
It is also likely that the compaction of ring conformations to less open conformations (Figure \ref{fig:eqlconf}(a)) may make it harder for rings to thread through each other.\cite{Smrek2019}
Nonetheless, it is clear that a substantial number of rings participate in self-threaded supramolecules, even for $\phi_R$ as low as 0.3 where each ring in the supramolecule is highly threaded by linear chains.
The dynamics of these objects is certainly complex and likely has a strong influence on the relaxation of these systems back to equilibrium.
This is a topic we will explore in future studies.

\section{CONCLUSIONS}

We have applied extensive molecular dynamics simulations to characterize the entanglement network topology, chain dynamics, and extensional rheology of symmetric ring-linear polymer blends with compositions systematically varied from pure linear to pure ring melts.
Linear and ring polymers with $Z=N/N_e\approx14$ entanglement segments perchain were blended to produce melts with a high degree of both entanglement and ring-linear threading.
As $\phi_R$ varies between pure linear to pure ring, a complex composite entanglement network forms which we directly characterize with primitive path analysis.
Linear chain conformations and diffusion times show little sensitivity to blending.
Linear primitive path lengths $L_{pp}^L$ do decrease as the ring fraction increases, but less than would be expected if rings acted as a dilutent because rings contribute additional topological constraints through ring-linear threading.
Ring polymers become highly threaded by linear chains as $\phi_R$ decreases, eventually embedding within a dense linear entanglement network as $\phi_R\rightarrow0$.
While neat rings have zero primitive path length $L_{pp}^R$, highly embedded rings have $L_{pp}^R$ comparable to linear chains. 
This  drives ring $R_g^R$ to swell by $\sim20\%$ as ring-linear threading increases.
Threading also slows ring diffusion by two orders of magnitude relative to neat rings, with the ring diffusion time $\tau_D^R\approx 2\tau_D^L$ for small $\phi_R$.

Simulations results for both the linear viscoelastic envelopes and nonlinear extensional stress growth of blends produce trends with $\phi_R$ and strain rate that agree with recent experiments.\cite{Borger2020,Peddireddy2020}
The coupling of ring and linear topologies produces a non monotonic increase in the zero-rate viscosity that peaks in the range $\phi_R=0.4-0.5$, in agreement with past simulations and experiments.\cite{Roovers1988,halverson12,Peddireddy2020}
Under uniaxial elongation flow blends develop a prominent overshoot in the stress $\sigma_E$ that we measure for different $\phi_R$ with the stress difference $\delta\sigma_E$ (Eq. \ref{eq:eq1}) between the blend startup and the ideal mixing of startup curves of the neat melts.
For sufficiently nonlinear strain rates, $\delta\sigma_E$ is largest for $\phi_R$ between $0.3-0.5$, and the peak occurs at a strain $\varepsilon_H\approx3.0$ that closely corresponds to the maximum extensibility of this model's entanglement segments $e^{\lambda_{\rm max}}\approx3.1$.

The nonlinear stress overshoot during extensional flow is driven by an overshoot and recoil of ring conformation stretch and orientation.
Meanwhile, linear conformations show relatively little little sensitivity to blending and elongate monotonically at all $\phi_R$.
Ring recoil is driven by the convective untrheading of rings from linear chains, as discussed by Borger \textit{et al.} and visualized for semidilute systems by Zhou and Young \textit{et al.}.\cite{Borger2020,Zhou2019,Young2020,Zhou2021}
Primitive path analyses directly visualize and quantify the topological changes of this unthreading transition and show the maximum of $\delta\sigma_E$ corresponds to the largest relative drops in $N_t$ and $L_{pp}^R$.
As $\phi_R\rightarrow1$, ring-linear threading decreases and $\sigma_E$ becomes dominated by a minority fraction of rings self-threading to form supramolecular daisy-chains.\cite{OConnor2020}
The fraction of rings participating in supramolecules $\phi_{\rm top}$ increases with $\phi_R$ and approaches a fraction $\sim0.2$ which was previously observed for pure rings at the same strain rate.\cite{OConnor2020} 

The maximum in the zero-shear viscosity and the nonlinear overshoot $\delta\sigma_E$ notably occur at similar $\phi_R$.
We believe this is not a coincidence and likely due to the existence of a maximum in the density of topological constraints in the composite entanglement network in this regime, before it breaks down as $\phi_R\rightarrow1$.
We presented simple arguments for why such a maximum should exist by considering the number of topological constraints contributed by threaded rings and entangled linear chains. 
This analysis estimated a maximum constraint density when the number of threadings contributed by rings equals the number of linear-linear entanglements formed by a linear chain $N_t=(1-\phi_R)Z$.
Direct measurements of $N_t$ (Figure \ref{fig:eqlstats}(d) estimate this maximum $\phi_R\approx0.4$, which corresponds closely with the onset of rapid drops in $L_{pp}^R$ and $\tau_d^R$ as $\phi_R$ (Figure \ref{fig:eqlstats}) increases and the entanglement network deteriorates (Figure \ref{fig:networks}).

We have shown that manipulating the composite entanglement topology of ring-linear enables \textit{physical} entangled melt processability without changing chemistry.
Simply changing $\phi_R$ allows the stress at the overshoot to be varied by nearly an order of magnitude for a modest nonlinear strain rate.
In addition, the peak stress at the maximum varies slowly for $\phi_R$ near 0.4, the estimated maximum in network constraint density.
This could mean that blend formulations seeking to maximize $\sigma_E$ --- in liquid or glassy states --- will be forgiving to variations in $\phi_R$.

Many interesting questions remain. We have focused on fluid rheology here, but it will be interesting to understand the mechanical behaviors of these blends in the glassy state. 
We are also intrigued by the dramatic topological reconfiguration of blends undergoing flow. 
The flows simultaneously unthreads ring-linear threadings and drives ring-ring self-threading, producing complex transient states with very different structure and dynamics from equilibrium.
It is likely that simple shear flows also convectively release ring-linear threadings, but they do not seem to siginficantly drive ring-ring supramolecular assembly due to the flow vorticity rotating chains.\cite{Jeong2020}
This fact may allow the prevalence of ring-linear and ring-ring threadings to be tuned in far-from-equilibrium blend states by applying mixed shear and extensional flows.
Regardless, the transient states after will likely produce long-lived and complex stress-relaxation behavior which we plan to explore in our future work.

\section{ACKNOWLEDGMENTS}
This work
was supported by the Sandia Laboratory Directed Research
and Development Program. This work was performed, in
part, at the Center for Integrated Nanotechnologies, an
Office of Science User Facility operated for the U.S.
Department of Energy (DOE) Office of Science. Sandia
National Laboratories is a multimission laboratory managed
and operated by National Technology \& Engineering
Solutions of Sandia, LLC, a wholly owned subsidiary of
Honeywell International, Inc., for the U.S. DOEs National
Nuclear Security Administration under Contract No.~DE-NA-
0003525. The views expressed in this article do not
necessarily represent the views of the U.S. DOE or the
United States Government.


%

\end{document}